\documentclass[journal]{IEEEtran}
\usepackage[latin1]{inputenc}
\usepackage{amsfonts}
\usepackage{amssymb}
\usepackage{graphicx}
\usepackage{verbatim}
\usepackage{array}
\usepackage{multirow}
\usepackage{dcolumn}
\usepackage{color}
\usepackage{flushend}
\usepackage{url}
\usepackage{stfloats}
\usepackage{siunitx}
\usepackage{rotating}
\usepackage{soul}
\usepackage{lscape}

\usepackage[tbtags]{amsmath}

\usepackage{eurosym}

\usepackage{multirow}

\usepackage{xcolor}
\usepackage[switch,pagewise]{lineno}

\newcommand{\bfg}[1]{\mbox{\boldmath $#1$\unboldmath}}

\newcommand{\fraca}[2]{\displaystyle\frac{#1}{#2}}

\def \R {{\rm I\kern -2.2pt R\hskip 1pt}}

\newcommand{\coleq}[1]{\mbox{\color{black}\boldmath $#1$\unboldmath \color{black}}}

\newcommand{\reviewcol}[1]{\color{black}#1 \color{black}}

\DeclareGraphicsExtensions{.png} 

\definecolor{grismedio}{gray}{0.3}

\definecolor{grisclaro}{gray}{0.7}

\newcolumntype{L}[1]{>{\raggedright\let\newline\\\arraybackslash\hspace{0pt}}m{#1}}
\newcolumntype{C}[1]{>{\centering\let\newline\\\arraybackslash\hspace{0pt}}m{#1}}
\newcolumntype{R}[1]{>{\raggedleft\let\newline\\\arraybackslash\hspace{0pt}}m{#1}}

\begin{document}
%%%\runningpagewiselinenumbers
%%%\linenumbers

\title{Robust Transmission Network Expansion Planning under Correlated Uncertainty}

\author{Cristina Rold\'an,
        Roberto M\'{\i}nguez,
        Raquel Garc\'{\i}a-Bertrand, \IEEEmembership{Senior Member, IEEE,}
        and Jos\'e M. Arroyo, \IEEEmembership{Senior Member, IEEE}

\thanks{This work was supported in part by the Ministry of Economy and Competitiveness of Spain under Project ENE2015-63879-R (MINECO/FEDER, UE).}

\thanks{C. Rold\'an, R. Garc\'{\i}a-Bertrand, and J. M. Arroyo are with the Departamento de Ingenier\'{\i}a El\'ectrica,
Electr\'onica, Autom\'atica y Comunicaciones, E.T.S.I. Industriales, Universidad
 de Castilla-La Mancha, Ciudad Real E-13071, Spain (e-mail: \mbox{cristinarg\_92@hotmail.com}; \mbox{Raquel.Garcia@uclm.es}; \mbox{JoseManuel.Arroyo@uclm.es}).

R. M\'{\i}nguez works as an independent consultant, Ciudad Real, Spain (e-mail: \mbox{rominsol@gmail.com}).

}}

% make the title area
\maketitle

\begin{abstract}
This paper addresses the transmission network expansion planning problem under uncertain demand and generation capacity. A two-stage adaptive robust optimization framework is adopted whereby the worst-case operating cost is accounted for under a given user-defined uncertainty set. This work differs from previously reported robust solutions in two respects. First, the typically disregarded correlation of uncertainty sources is explicitly considered through an ellipsoidal uncertainty set relying on their variance-covariance matrix.
\reviewcol{In addition, we describe the analogy between the corresponding second-stage problem and a certain class of mathematical programs arising in structural reliability. This analogy gives rise to a relevant probabilistic interpretation of the second stage, thereby revealing an undisclosed feature of the worst-case setting characterizing robust optimization with ellipsoidal uncertainty sets.}
 More importantly, a novel nested decomposition approach based on results from structural reliability is devised to solve the proposed robust counterpart, which is cast as an instance of mixed-integer trilevel programming. \color{black}Numerical results from several case studies demonstrate that the effect of correlated uncertainty can be captured by the proposed robust approach.\color{black}
\end{abstract}

%\vspace{-0.3cm}

\section*{Nomenclature}

This section lists the main notation used throughout the paper. Symbol ``\hspace{5pt}$\hat{ }$\hspace{5pt}'' is used to represent the optimal value of a specific variable. Additional symbols with superscripts ``$(k)$'' and ``$(m)$'' are used to indicate the value of a specific variable at iterations $k$ and $m$ of the outer loop, respectively. Similarly, superscript ``$(\nu)$'' is used to denote results obtained at iteration $\nu$ of the inner loop.
%\vspace{-0.3cm}

\subsection{Sets}
\begin{description}
[\IEEEsetlabelwidth{$\mbox{\it{ELLLLL}}$}\IEEEusemathlabelsep]

\item[$\Omega(\cdot)$] Feasibility set of {\bfg x}.
\item[${\bfg D}$] Uncertainty set.

\end{description}
%\vspace{-0.3cm}

\subsection{Constants}
\begin{description}
[\IEEEsetlabelwidth{$\mbox{\it{ELLLLL}}$}\IEEEusemathlabelsep]

\item[${\beta}$] Conservativeness parameter.

\item[$\epsilon^{\rm {IL}}$] Convergence tolerance for the inner loop.

\item[$\epsilon^{\rm {OL}}$] Convergence tolerance for the outer loop.

\item[$\Pi$] Investment budget.

\item[$\sigma$] Scaling factor.

\item[${\bfg \Sigma}$] Variance-covariance matrix.

\item[${\bfg b}$] Vector of operating cost coefficients.

\item[${{\bfg C^{\rm{L}}}}$] Vector of investment cost coefficients.

\item[$\bar {\bfg d}$] Vector of mean or forecast values.

\item[${\bfg L}$] Matrix used in the affine mapping of ${\bfg d}$.

\end{description}

%\vspace{-0.3cm}

\subsection{Decision Variables}

\begin{description}
[\IEEEsetlabelwidth{$\mbox{\it{ELLLLL}}$}\IEEEusemathlabelsep]

\item[$\alpha$] Approximation of the worst-case operating cost.

\item[${\bfg d}$] Vector of uncertain generation capacities and demands.

\item[${\bfg d}^{\rm D}$] Vector of uncertain demands.

\item[${\bfg d}^{\rm G}$] Vector of uncertain generation capacities.

\item[${\bfg v}$] Vector of binary variables representing expansion decisions.

\item[${\bfg x}$] Vector of second-stage variables associated with the system operation under uncertainty such as production and consumption levels, load shedding, and network-related variables.

\item[${\bfg {x}_m}$] Vector of operation-related variables under the worst case identified at iteration $m$.

\item[${\coleq y}$] \color{black}Vector of uncertain generation capacities and demands used for the analogy with structural reliability methods.\color{black}

\end{description}
%\vspace{-0.3cm}

\subsection{Functions}
\begin{description}
[\IEEEsetlabelwidth{$\mbox{\it{ELLLLL}}$}\IEEEusemathlabelsep]

\item[$\Phi(\cdot)$] Cumulative distribution function of the standard normal distribution.
\item[$c^{{\rm O},min}(\cdot)$] Minimum operating cost.
\item[$c^{{\rm O},wc}(\cdot)$] Worst-case operating cost.
\item[${f_{\rm obj}}(\cdot)$] \reviewcol{Objective function of the analogous problem.}

\end{description}
%\vspace{-0.3cm}

\subsection{Others}
\begin{description}
[\IEEEsetlabelwidth{$\mbox{\it{ELLLLL}}$}\IEEEusemathlabelsep]

\item[${\bfg \eta}$] Vector of dual variables associated with the constraints fixing the values of ${\bfg d}$.

\item[$\mu$] \reviewcol{Dual variable of the analogous problem.}

\item[$C^{\rm MP}$] Approximation of the optimal total cost provided by the master problem.

\item[$C^{\rm SP}$] Approximation of the optimal total cost provided by the subproblem.

\item[${\bfg z}$] Vector of perturbations associated with the affine mapping of ${\bfg d}$.
\end{description}

%\vspace{-0.3cm}

\section{Introduction}

\color{black}
\IEEEPARstart{D}{ecision} making associated with the investment in the transmission network plays a key role in power system planning in both centralized and competitive frameworks. This paper addresses the transmission network expansion planning problem under correlated uncertainty. Transmission network expansion planning consists in determining how to expand and reinforce the transmission network in order to meet the future demand with the available generation assets \cite{LumbrerasR:16}. The thrust of this paper is the integration of the two main uncertainty sources characterizing this planning problem, namely demand and available generation capacity. Uncertain demand growth has been the primary driver of conventional transmission network expansion planning models. Within the current context where the growing installation of renewable-based generation assets comes into play, network planners also face unprecedented levels of uncertainty related to available generation capacity. Moreover, due to their strong relation to meteorological phenomena, the spatial correlations of renewable energy sources and demand consumptions within a given geographical area may be significant \cite{Morales:10}--\cite{BellWFH:15}.
% \cite{Morales:10}, \cite{XieCFLKPI:11}, \cite{BellWFH:15}.
 Thus, this paper is devoted to accounting for the effect of such correlated uncertainty on transmission network investment decisions.
\color{black}

%%%Network planners are responsible for determining how to expand and reinforce the transmission network in order to meet the future
%%%demand with the available generation assets \cite{LumbrerasR:16}. To that end, transmission network expansion planning models have been traditionally implemented while considering to some extent uncertainty, which was primarily related to demand growth. Within the current context where the growing installation of renewable-based generation assets comes into play, unprecedented levels of uncertainty are faced by network planners. Moreover, due to their strong relation to meteorological phenomena, the spatial correlation of renewable energy sources within a given geographical area may be significant \cite{Morales:10}. As a consequence, traditional transmission network expansion planning models have to be adapted accordingly.

Researchers have begun to examine two-stage adaptive robust optimization \cite{Ben-Tal:04} as a suitable framework to handle uncertainty in the transmission network expansion planning problem \cite{WuCX:08}--\cite{ZhangC:17}.
%%%\cite{WuCX:08}--\cite{Jabr:13}--\cite{ChenWWHW:14}--\cite{Ruiz:15}--\cite{MinguezG:16}--
%%%\cite{MinguezGAA:17}--\cite{ZhangC:17}.
\color{black}
Wu \emph{et al.} \cite{WuCX:08} applied adjustable robust optimization to the transmission network expansion planning problem under uncertain nodal demands. The resulting trilevel program was addressed by a greedy randomized adaptive search procedure relying on the duality-based transformation of the second-stage max-min problem to a nonlinear single-level equivalent, which was solved by a modified branch-and-bound algorithm.
Jabr \cite{Jabr:13} extended the model described in \cite{WuCX:08} to include uncertain renewable-based generation and a cardinality uncertainty set where uncertainty budgets were introduced to control the conservativeness of the fluctuation-interval-based uncertainty characterization. Jabr proposed the application of Benders decomposition, which involved the iterative solution of a master problem and a max-min subproblem associated with the second stage. The original max-min subproblem was cast as a single-level equivalent relying on mixed-integer linear programming.
Chen \emph{et al.} \cite{ChenWWHW:14} analyzed the same problem presented in \cite{Jabr:13} using a polyhedral uncertainty set and two different criteria, namely minimax cost and minimax regret. The resulting trilevel robust counterparts were addressed by the column-and-constraint generation algorithm \cite{ZengZ:13}, thereby also involving the iterative solution of a master problem and a max-min subproblem associated with the second stage. The max-min subproblem was transformed into an equivalent mixed-integer linear program using Karush-Kuhn-Tucker optimality conditions.
Ruiz and Conejo \cite{Ruiz:15} addressed essentially the same planning model analyzed in \cite{Jabr:13} and \cite{ChenWWHW:14} with the methodology described in \cite{ChenWWHW:14}, albeit for a different polyhedral uncertainty set.
M\'{\i}nguez and Garc\'{\i}a-Bertrand \cite{MinguezG:16} proposed a column-and-constraint generation algorithm relying on the duality-based single-level transformation of the resulting max-min subproblem to effectively solve the robust counterparts modeled in \cite{Jabr:13}--\cite{Ruiz:15} for the cardinality uncertainty set used in \cite{Jabr:13}.
 Using the formulation and the column-and-constraint generation algorithm presented in \cite{MinguezG:16},  M\'{\i}nguez \emph{et al.} \cite{MinguezGAA:17} developed a block coordinate descent method for the subproblem, which was suitable for practical instances.
Finally, Zhang and Conejo \cite{ZhangC:17} described a robust planning model incorporating both long- and short-term uncertainties. In \cite{ZhangC:17}, the solution approach relied on a column-and-constraint generation algorithm wherein the resulting bilinear subproblem associated with the second stage was solved by an outer approximation.

In \cite{WuCX:08}--\cite{ZhangC:17}, \color{black} the optimal expansion plan considers the worst-case realization of uncertain parameters in a user-defined set, referred to as uncertainty set. Thus, uncertain parameters representing uncertain nodal demands and generation capacities are modeled as decision variables. Such an endogenous characterization of uncertainty gives rise to a robust counterpart that is formulated as a mixed-integer trilevel program where first-stage decisions are associated with the upper-level problem whereas the second stage corresponds to the two lowermost optimizations.
 The upper level determines the least-cost investment decisions while considering the operation under all uncertainty realizations within a pre-specified uncertainty set. For a given upper-level decision vector, the middle level identifies the worst-case parameter values maximizing the operating cost. Finally, the lower level models the least-cost system operation for the investment decisions and the uncertainty realizations determined in the upper and middle level, respectively. \color{black}
Unfortunately, available robust models for transmission network expansion planning \cite{WuCX:08}--\cite{ZhangC:17} %%%\cite{WuCX:08}--\cite{Jabr:13}--\cite{ChenWWHW:14}--\cite{Ruiz:15}--\cite{MinguezG:16}--
%%%\cite{MinguezGAA:17}--\cite{ZhangC:17}
rely on cardinality and polyhedral uncertainty sets disregarding the correlation of uncertain parameters. It is worth mentioning that the correlation of wind power generation \cite{Morales:10}, \cite{XieCFLKPI:11} and demand \cite{BellWFH:15} may play a crucial role when investment decisions are involved.\color{black}

Motivated by the modeling limitation characterizing previously reported robust models \cite{WuCX:08}--\cite{ZhangC:17}, this paper is focused on the proposal of a novel two-stage adaptive robust approach for transmission network expansion planning where correlated uncertainty sources are explicitly accounted for. As a salient modeling aspect, rather than using previous cardinality and polyhedral uncertainty sets, the proposed approach relies on an ellipsoidal uncertainty set based on the first and second moments of the probability distribution of the uncertain parameters, i.e., mean or forecast values and the variance-covariance matrix \reviewcol{\cite{Ben-TalN:98}--\cite{Ben-TalN:00}.}

The consideration of an ellipsoidal uncertainty set gives rise to two major modifications with respect to the trilevel robust counterparts addressed in \cite{WuCX:08}--\cite{ZhangC:17}: 1) nonlinear terms are part of the middle-level problem, and 2) an equivalent discrete representation of the uncertainty set can no longer be obtained. The solution to this type of trilevel programs constitutes a challenging field that is still unsolved by the operations research community. Note that the unavailability of a binary-variable-based equivalent for uncertain demands and generation capacities precludes the application of the solution approaches previously reported in \cite{WuCX:08}--\cite{ZhangC:17}.

\enlargethispage*{0.5cm}

As a salient methodological feature, this paper proposes a novel nested decomposition approach involving two loops. In the outer loop, the original problem is decomposed into a master problem and a subproblem through a column-and-constraint generation algorithm \cite{ZengZ:13}. The inner loop is devoted to the solution of the max-min subproblem at each iteration of the outer loop. \reviewcol{Due to the use of an ellipsoidal uncertainty set, the second-stage problem and hence the max-min subproblem are analogous to an instance of mathematical programming arising in structural reliability, for which a decomposition technique was developed in \cite{MinguezCG:11}.}
Using this analogy, the inner loop comprises the application of such a decomposition method to the max-min subproblem.

Similar to previous works \cite{WuCX:08}--\cite{ZhangC:17}, global optimality is not guaranteed and a measure of the distance to the global optimum is not provided. This drawback is overcome here by restarting the inner loop so that most of the solution space is searched. The proposed multi-start decomposition framework allows avoiding local optima, eventually reaching the global optimum. Moreover, the above analogy allows straightforwardly assessing the quality of the solution of the inner loop through an out-of-sample assessment based on Monte Carlo simulation. Thus, the proposed technique represents a valuable approximation in the absence of exact solution methodologies.

\color{black}Admittedly, alternative uncertainty sets, such as the polyhedral instances applied in \cite{GuanW:14}--\cite{LorcaS:17} %\cite{GuanW:14}--\cite{MoreiraSA:15}--\cite{LiGWZ:15}--\cite{LiZZL:16}--\cite{LorcaS:15}--\cite{LorcaS:17} %
for robust generation scheduling and dispatch, can be used to incorporate correlation in robust transmission network expansion planning. Bearing in mind that, to the best of the authors' knowledge, this is the first attempt to address robust transmission network expansion planning under correlated uncertainty, the adoption of an ellipsoidal rather than polyhedral uncertainty set is motivated by the following three reasons:
\reviewcol{
\begin{enumerate}
  \item The correlation structure of uncertain parameters is considered through their variance-covariance matrix, which can be obtained in a straightforward and simple way.
  \item The level of conservativeness is easily controlled by a single parameter, which is beneficial for practical implementation purposes.
  \item Relevant and methodologically advantageous findings of structural reliability are leveraged. Such findings enable:
    \begin{enumerate}
        \item The development of a computationally efficient and scalable solution approach, wherein one of the optimization problems that are iteratively solved has an analytical solution.
        \item The informed choice for the conservativeness parameter.
        \item The validation of the solution quality based on an out-of-sample assessment relying on Monte Carlo simulation.
    \end{enumerate}
\end{enumerate}
}

Such motivations are backed by the numerical experience reported in this paper, where we show the performance of the proposed approach with several case studies including a practical benchmark based on the Polish 2383-bus system.\color{black}

The major contributions of this paper are as follows:
 \begin{enumerate}
   \item \reviewcol{We propose a novel two-stage adaptive robust optimization model with an ellipsoidal uncertainty set for transmission network expansion planning that allows incorporating correlated uncertainty.}
   \item \color{black}Based on our numerical experience, we show that disregarding the effect of correlated uncertainty may lead to suboptimality. \color{black}
   \item  \reviewcol{We describe the relevant analogy between the second stage and a certain class of mathematical programs that are of interest within the context of structural reliability. This analogy allows quantifying the solution quality.}
   \item We present a novel nested approach combining the column-and-constraint generation algorithm and a decomposition-based method previously applied to structural reliability analysis. \color{black}Note that the structural reliability problems upon which our work is built involve a single optimization level. Thus, the proposed approach represents an original extension of the application scope of structural reliability methods to a multilevel decision-making framework. \color{black}
 \end{enumerate}

The remainder of the paper is structured as follows. Section~\ref{rtnep} presents the proposed two-stage robust planning model.
\reviewcol{Section~\ref{s.analogy} describes the analogy between the second-stage problem and an instance of mathematical programming used in structural reliability.}
Section~\ref{s22} is devoted to the solution approach. In Section~\ref{CaseStudy}, numerical results are provided and analyzed. Relevant conclusions are drawn in Section~\ref{Conclusions}. Finally, the Appendix provides a detailed formulation  of the lower-level problem.

\vspace{-0.2cm}
\section{Proposed Model}\label{rtnep}

Using a compact formulation, this section presents an ellipsoidal uncertainty set suitable for the correlation of uncertainty sources followed by the proposed robust counterpart for transmission network expansion planning under uncertainty.

\vspace{-0.3cm}

\subsection{Ellipsoidal Uncertainty Set}

Network planners face two main sources of uncertainty, namely generation capacities ${\bfg d}^{\rm G}$ and demands ${\bfg d}^{\rm D}$, which are hereinafter represented by the uncertain parameter vector ${\bfg d}=({\bfg d}^{\rm G},\; {\bfg d}^{\rm D})^T$. Based on \cite{Ben-TalN:99}, we propose modeling the correlated uncertainty of such parameters through an ellipsoidal uncertainty set, denoted by ${\bfg D}$, which can be expressed using the Mahalanobis distance as follows:
\begin{align}
&{\bfg D}=\left\{{\bfg d}:\left({\bfg d}-\bar{\bfg d}\right)^T \left({\bfg \Sigma}\right)^{-1}\left({\bfg d}-\bar{\bfg d}\right)\le \beta^2 \right\}.\label{eq.Mahalanobis}%\\
\end{align}

%%%For simplicity, the uncertainty set (\ref{eq.Mahalanobis}) is transformed as follows using an affine mapping into a ball of radius $\beta$ \cite{Ben-TalN:99}:
%%%%
%%%\begin{align}
%%%&{\bfg d}=\bar {\bfg d}+ {\bfg L}{\bfg z}\label{eq.inconsRCellip2}\\
%%%&\|{\bfg z}\|\le \beta\label{eq.inconsRCellip22}%\\
%%%\end{align}
%%%%
%%%\noindent where ${\bfg L}$ is the mapping matrix, which can be obtained from the Cholesky factorization of the variance-covariance matrix ${\bfg \Sigma}={\bfg L}{\bfg L}^{T}$; ${\bfg z}$ represents a perturbation vector; and $\|\cdot\|$ stands for the Euclidean norm.

The ellipsoidal uncertainty set formulated in (\ref{eq.Mahalanobis})
%(or  (\ref{eq.inconsRCellip2}) and (\ref{eq.inconsRCellip22}), likewise)
features two advantages over the cardinality and polyhedral uncertainty sets used in existing robust counterparts for transmission expansion planning \cite{WuCX:08}--\cite{ZhangC:17}, namely 1) correlation is precisely accounted for through the variance-covariance matrix ${\bfg \Sigma}$, thereby allowing a more realistic characterization of uncertainty, and 2) a single nonnegative conservativeness parameter, $\beta$, is considered, thereby enabling a simpler control of the degree of robustness.

\vspace{-0.3cm}

\subsection{Compact Problem Formulation}

\color{black}Under the worst-case setting featured by adjustable robust optimization \cite{Ben-Tal:04}, the proposed robust expansion planning model determines the optimal network investment decisions in the first stage, i.e., before the realization of uncertainty, while accounting for the optimal reaction against uncertainty in the second stage. The proposed two-stage robust counterpart can be cast as the following trilevel program:
\begin{align}
&\underset{{\bfg v}}{\text{Minimize}} \quad{{{\bfg C^{\rm{L}}}}^T{\bfg v}+\sigma c^{{\rm O},wc}({\bfg v})}\label{eq1}\\
&\text{subject to:}\notag\\
& {{\bfg C^{\rm{L}}}}^T{\bfg v}  \le  \Pi \label{eq2}\\
&{\bfg v} \in \{0,1\} \label{eq3}
\end{align}
where $c^{{\rm O},wc}({\bfg v})$ is provided by:
\begin{align}
&c^{{\rm O},wc}({\bfg v})=\underset{{\bfg d}}{\text{Maximize}} \quad c^{{\rm O},min}({\bfg v},{\bfg d})\label{eq4}\\
&\qquad\qquad\quad\text{subject to:}\notag\\
%&\qquad\qquad\quad \text{Constraints (\ref{eq.inconsRCellip2})--(\ref{eq.inconsRCellip22})} \label{eq5}%\\
%&\qquad\qquad\quad \text{Constraint (\ref{eq.Mahalanobis})} \label{eq5}%\\
&\qquad\qquad\quad\left({\bfg d}-\bar{\bfg d}\right)^T \left({\bfg \Sigma}\right)^{-1}\left({\bfg d}-\bar{\bfg d}\right)\le \beta^2\label{eq5}
\end{align}
\noindent and where $c^{{\rm O},min}({\bfg v},{\bfg d})$ is obtained from:
%\vspace{-0.1cm}
\begin{align}
&c^{{\rm O},min}({\bfg v},{\bfg d})=\underset{{\bfg x}}{\text{Minimize}} \quad {{\bfg b}^T}{\bfg x}\label{eq6}\\
&\qquad\qquad\qquad\hspace{0.2cm}\text{subject to:}\notag\\
&\qquad\qquad\qquad\hspace{0.2cm} {\bfg x}\in \Omega({\bfg v},{\bfg d}). \label{eq7}%\\
\end{align}

Note that the upper-level problem (\ref{eq1})--(\ref{eq3}) represents the first stage, whereas the two lowermost optimization levels  (\ref{eq4})--(\ref{eq7}) model the second stage.\color{black}

The upper-level problem (\ref{eq1})--(\ref{eq3}) determines the optimal expansion plan $\bfg v$ minimizing a composite objective function comprising the investment cost and the worst-case operating cost subject to an investment budget (\ref{eq2}) and the integrality of $\bfg v$ (\ref{eq3}). The scaling factor $\sigma$ is used to make investment and worst-case operating costs comparable quantities.

%%%\color{black}, and it is equal to the number of hours in one year 8760 divided by capital recovery factor $R$. Note that $R\left({\bfg C^{\rm{L}}}^T{\bfg v}\right)$ correspond to annualized investment costs.\color{black}

The second-stage max-min problem (\ref{eq4})--(\ref{eq7}) provides the worst-case operating cost, $c^{{\rm O},wc}({\bfg v})$.
In the middle-level problem (\ref{eq4})--(\ref{eq5}), the maximization of the minimum operating cost in (\ref{eq4}) identifies the vector of worst-case uncertainty realizations $\bfg d$ within the ellipsoidal uncertainty set (\ref{eq5}). In turn, the lower-level problem (\ref{eq6})--(\ref{eq7})  provides the vector of operating decisions $\bfg x$ minimizing the operating cost (\ref{eq6}) over the feasibility space (\ref{eq7}) for given vectors $\bfg v$ and $\bfg d$. As is customary in the related literature \cite{WuCX:08}--\cite{ZhangC:17}, the lower-level problem (\ref{eq6})--(\ref{eq7}) is a linear program parameterized in terms of $\bfg v$ and $\bfg d$. For quick reference, a detailed formulation of the lower-level problem (\ref{eq6})--(\ref{eq7}) can be found in the Appendix.

%\hl{Note that the second-stage problem ({\ref{eq4}})--({\ref{eq7}}) comprises the maximization of an objective function ({\ref{eq4}}) dependent on the decision vector {${\bfg d}$}, which is solely constrained by ({\ref{eq5}}). Thus, at the optimum, constraint ({\ref{eq5}}) holds as an equality.}

\color{black}An important feature of problem ({\ref{eq4}})--({\ref{eq7}}) is that, as per ({\ref{eq6}})--({\ref{eq7}}), {$c^{{\rm O},min}({\bfg v},{\bfg d})$} is a monotonically increasing function of {${\bfg d}^{\rm D}$} and a nonincreasing function of {${\bfg d}^{\rm G}$}. As a consequence, as expression ({\ref{eq5}}) represents a boundary for {${\bfg d}$}, such a constraint holds as an equality at the optimum.\color{black}

%This can be verified by noticing that the worst realization of ``nature'' would seek, for a fixed network configuration, to cause the maximum level of load shedding (and thus the maximum operating costs). Therefore, it will try to make the generation capacity to be as lower as possible and$/$or the demand load to be as higher as possible until reaching the boundary of the uncertainty set.

%%%Due to the bilevel structure of the second-stage problem (\ref{eq4})--(\ref{eq7}), the proposed robust counterpart for transmission network expansion planning (\ref{eq1})--(\ref{eq3}) is an instance of mixed-integer trilevel programming.

\section{\color{black}Analogy between the Second-Stage Problem and Structural Reliability Analysis\color{black}}\label{s.analogy}

\reviewcol{A relevant analogy can be drawn between the second-stage problem (\ref{eq4})--(\ref{eq7}) and the following mathematical program:}
\begin{align}
&\underset{{\coleq y}}{\text{Minimize}} \quad
{f_{\rm obj}}({\coleq y})=\left({\coleq y}-\bar{\bfg d}\right)^T \left({\bfg \Sigma}\right)^{-1}\left({\coleq y}-\bar{\bfg d}\right)
\label{eqa1}\\
&\text{subject to:}\notag\\
&c^{{\rm O},min}({\coleq v},{\coleq y}) \geq c^{{\rm O},wc}({\bfg v}): {\mu} \label{eqa2}
\end{align}
\noindent \reviewcol{where ${\bfg y}$ denotes a vector of decision variables modeling uncertain parameters, $c^{{\rm O},wc}({\bfg v})$ represents the optimal value of the objective function maximized in ({\ref{eq4}}), $\mu$  is the dual variable associated with ({\ref{eqa2}}), and $c^{{\rm O},min}({\bfg v},{\bfg y})$ is obtained from:}
%\vspace{-0.1cm}
\begin{align}
&c^{{\rm O},min}({\coleq v},{\coleq y})=\underset{{\bfg x}}{\text{Minimize}} \quad {{\bfg b}^T}{\bfg x}\label{eq6y}\\
&\qquad\qquad\qquad\hspace{0.15cm}\text{subject to:}\notag\\
&\qquad\qquad\qquad\hspace{0.15cm} {\bfg x}\in \Omega({\coleq v},{\coleq y}). \label{eq7y}%\\
\end{align}

\reviewcol{Expressions (\ref{eqa1})--(\ref{eqa2}) correspond to the formulation associated with the structural reliability First-Order-Second-Moment approach \cite{HasoferL:74}. This method only uses the first two moments of the probability distribution of the uncertain parameters (mean and variance-covariance), which is equivalent to assuming that uncertain parameters are normally distributed. The First-Order-Second-Moment approach can be used to estimate the probability of obtaining an operating cost not exceeding $c^{{\rm O},wc}({\bfg v})$.}
Hereinafter, this probability is referred to as confidence level. Bearing in mind that, in general, the minimum operating cost $c^{{\rm O},min}({\coleq v},{\coleq y})$ is a piecewise linear function with respect to vector ${\coleq y}$, the confidence level can be estimated approximately as \cite{HasoferL:74}:
\vspace{-0.2cm}
 \begin{equation}\label{Probapprox}
   \textrm{Prob}\bigl(c^{{\rm O},min}({\coleq v},{\coleq y})\leq c^{{\rm O},wc}({\bfg v})\bigr) \cong \Phi\Bigl(\sqrt{f_{\rm obj}(\hat{\coleq y})}\Bigr)
 \end{equation}
%\vspace{-0.2cm}
\noindent where {${\hat{\coleq y}}$} is the optimal solution to problem ({\ref{eqa1}})--({\ref{eq7y}}) and, as per ({\ref{eqa1}}), the optimal value of the objective function is $f_{\rm obj}(\hat{\coleq y})=\left({\hat{\coleq y}}-\bar{\bfg d}\right)^T \left({\bfg \Sigma}\right)^{-1}\left({\hat{\coleq{ y}}}-\bar{\bfg d}\right)$.

\begin{figure}[t!]
  \begin{center}
  \includegraphics[width=0.42\textwidth]{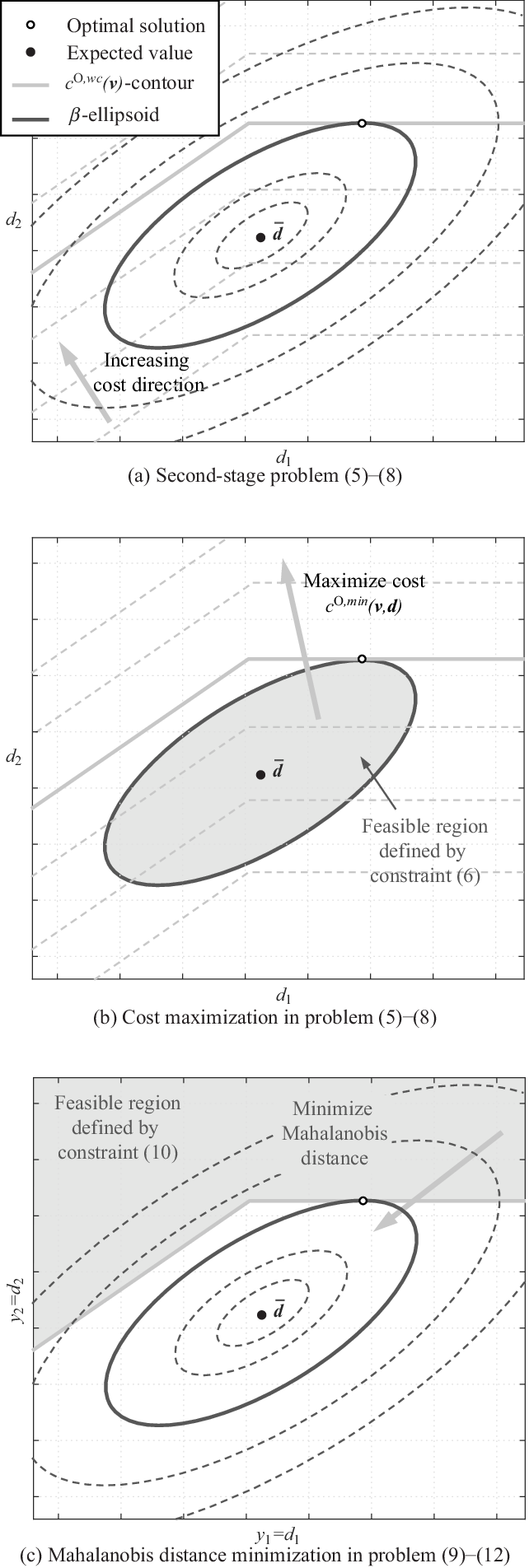}
%    \caption{\reviewcol{Graphical interpretation of the analogy between the second-stage problem (\ref{eq4})--(\ref{eq7}) and problem (\ref{eqa1})--(\ref{eq7y}).}}
    \caption{\reviewcol{Graphical interpretation of the analogy.}}
    \label{analogy}
  \end{center}
%\vspace{-0.5cm}
\end{figure}

\reviewcol{As explained in \cite{HasoferL:74}, the First-Order-Second-Moment method provides an approximation of the confidence level due to the simplification adopted in the probability calculation and the use of only the first two moments of the probability distribution of the uncertain parameters.}
\color{black}Although (\ref{Probapprox}) constitutes an approximation, it is a good estimate which improves its accuracy as the value of $\sqrt{f_{\rm obj}(\hat{\bfg y})}$ increases, as is shown in the computational experiments.

Analogously to problem (\ref{eq4})--(\ref{eq7}), constraint ({\ref{eqa2}}) holds as an equality at the optimal solution to problem ({\ref{eqa1}})--({\ref{eq7y}}), {$\hat {\bfg y}$}. Thus, in case no degeneracy exists for $c^{{\rm O},min}({\bfg v},\hat{\bfg y})$, {$\hat {\bfg y}$} must satisfy the following Karush-Kuhn-Tucker necessary optimality conditions:\color{black}
%
%\newpage
\begin{align}
&\left.{\bfg \nabla}_{\coleq y}\!\!\left\{\left({\coleq y}-\bar{\bfg d}\right)^T \!\!{\bfg \Sigma}^{-1}\!\!\left({\coleq y}-\bar{\bfg d}\right)\right\}\!\!\right|_{\hat {\coleq y}}\!\!+\!\!\mu\!\left.{\bfg \nabla}_{\bfg y} \left\{
c^{{\rm O},min}({\coleq v},{\coleq y})\right\}\right|_{\hat {\coleq y}}\!=\!0\label{KKT5}\\
&c^{{\rm O},min}({\coleq v},\hat{\coleq y})-c^{{\rm O},wc}({\bfg v})=0\label{KKT6}
%\\
%&\mu<0,\label{KKT7}
\end{align}
\color{black}
\noindent where we are taking into consideration that the conservativeness parameter $\beta$ is positive. It is worth mentioning that $\beta=0$ represents the deterministic case for which conditions (\ref{KKT5}) and (\ref{KKT6}) do not hold. In this case, the only feasible solution corresponds to the expected values $\hat {\bfg y}=\bar{\bfg d}$, which is not of interest within a robust framework.
\color{black}

Note that (\ref{KKT5}) can be equivalently cast as:
\begin{align}
&\fraca{1}{\mu}\!\!\left.{\bfg \nabla}_{\coleq y}\!\!\left\{\left({\coleq y}-\bar{\bfg d}\right)^T \!\!{\bfg \Sigma}^{-1}\!\!\left({\coleq y}-\bar{\bfg d}\right)\!\!\right\}\right|_{\hat {\coleq y}}\!\!+\!\!\left.{\bfg \nabla}_{\bfg y} \left\{
c^{{\rm O},min}({\coleq v},{\coleq y})\right\}\right|_{\hat {\coleq y}}=0.\label{KKT1}
%\\
%&\fraca{1}{\mu}<0.\label{KKT3}
\end{align}

Moreover, from (\ref{eqa1}):
\begin{equation}\label{KKT4}
\left({\hat{\coleq y}}-\bar{\bfg d}\right)^T \left({\bfg \Sigma}\right)^{-1}\left({\hat{\coleq{ y}}}-\bar{\bfg d}\right)-f_{\rm obj}(\hat{\coleq y})=0.
\end{equation}

\color{black}According to {\cite{CastilloCMC:06}}, \color{black}expressions ({\ref{KKT1}}) and ({\ref{KKT4}}) correspond to the Karush-Kuhn-Tucker necessary optimality conditions for the middle-level problem (\ref{eq4})--(\ref{eq5}) if the following conditions hold:
\begin{enumerate}
  \item \reviewcol{The vector of decision variables ${\bfg y}$ becomes ${\bfg d}$ so that $\hat{\bfg y}=\hat{\bfg d}$.}
  \item The gradient of the minimum operating cost $c^{{\rm O},min}({\bfg v},{\bfg y})$ with respect to ${\bfg y}$ particularized at $\hat {\bfg y}$, i.e., $\left.{\bfg \nabla}_{\bfg y} \left\{
c^{{\rm O},min}({\bfg v},{\bfg y})\right\}\right|_{\hat {\bfg y}}$ is unique so that no degeneracy exists.
  \item $\beta^2=f_{\rm obj}(\hat{\bfg y})=f_{\rm obj}(\hat{\bfg d})$.
\end{enumerate}

In such a case, \color{black} using (\ref{Probapprox}),  $\Phi(\beta)$ represents an approximate estimation for the confidence level associated with the optimal solution to (\ref{eq4})--(\ref{eq7}), i.e., $\textrm{Prob}(c^{{\rm O},min}({\bfg v},{\bfg d})\leq c^{{\rm O},wc}({\bfg v}))$.

\reviewcol{The analogy between the second-stage problem and structural reliability analysis is graphically illustrated in Fig.~\ref{analogy}
for a two-dimensional example. In the three subplots of Fig.~\ref{analogy}, the solid gray line represents the contour of the optimal cost $c^{{\rm O},wc}({\bfg v})$ in (\ref{eq4}) and (\ref{eqa2}), whereas the solid black line represents the ellipsoid with Mahalanobis distance equal to $\beta$ defined by constraint (\ref{eq5}). Analogously, dashed gray and black lines represent other cost contours and ellipsoids, respectively. In Fig.~\ref{analogy}(a), the optimal solution to problem (\ref{eq4})--(\ref{eq7}) is represented in the space of uncertain parameters ${\bfg d}$. Fig.~\ref{analogy}(b) shows how the optimal solution to problem (\ref{eq4})--(\ref{eq7}) is achieved by maximizing $c^{{\rm O},min}({\bfg v},{\bfg d})$ within the feasible region (gray shaded area) defined by the ellipsoid associated with (\ref{eq5}). As can be observed in Fig.~\ref{analogy}(c), the same solution can be attained by minimizing the Mahalanobis distance in (\ref{eqa1}) within the feasible region (gray shaded area) defined by constraint (\ref{eqa2}).}

The relevance of the above analogy for the solution of the original problem (\ref{eq1})--(\ref{eq7}) is twofold. \color{black}First, we can benefit from the availability of a solution procedure, as described in Section \ref{s22}. \color{black} In addition, the solution to the second stage represents an approximation of the value-at-risk (or quantile) of the minimum operating cost associated with a confidence level given by $\Phi(\beta)$. Such a probabilistic interpretation is useful for the network planner in order to make an informed decision on the choice for the conservativeness parameter $\beta$. Moreover, the solution quality of the second stage can be straightforwardly validated by an out-of-sample assessment based on Monte Carlo simulation.  \color{black}Due to the use of the approximate expression (\ref{Probapprox}), the confidence level resulting from the out-of-sample assessment may differ from the estimation provided in (\ref{Probapprox}). Thus, the difference represents an indicator of the quality of the approximation and the appropriateness of the selected $\beta$.\color{black}

\section{Solution Approach}\label{s22}

This section describes the novel two-loop approach proposed to address problem (\ref{eq1})--(\ref{eq7}). The outer loop consists in the application of the column-and-constraint generation algorithm to the original trilevel robust counterpart, thereby giving rise to the iterative solution of a single-level master problem and a max-min subproblem. The inner loop is another iterative procedure whereby the max-min subproblem at each outer-loop iteration is approximately solved by the decomposition-based method described in \cite{MinguezCG:11}.

\vspace{-0.5cm}
\subsection{Master Problem}
The master problem constitutes a relaxation for problem (\ref{eq1})--(\ref{eq7}) where a set of valid operating constraints are iteratively added. The addition of such constraints, which are set up with information from the subproblem, allows obtaining a more robust expansion plan at each iteration. At iteration $k$ of the column-and-constraint generation algorithm, the master problem is formulated as the following mixed-integer linear program:
\vspace{-0.2cm}
\begin{align}
&\underset{\alpha,{\bfg v},{\bfg {x}_m}}{\text{Minimize}} \quad{{{\bfg C^{\rm{L}}}}^T{\bfg v}+\sigma \alpha}\label{mof1}\\
&\text{subject to:}\notag\\
&\text{Constraints (\ref{eq2})--(\ref{eq3})}\label{mcompact}\\
&\alpha \geq {{\bfg b}^T}{\bfg {x}_m}; m=1,\ldots, k-1  \label{mof3}\\
&{\bfg {x}_m}\in \Omega({\bfg v},{\bfg {d}^{(m)}}); m=1,\ldots, k-1\label{mof4}\\
&\alpha  \geq 0 \label{neg}%\\
\end{align}
\noindent where the additional vectors of decision variables ${\bfg {x}_m}$ corresponding to ${\bfg {x}}$ are associated with the uncertainty realizations identified by the subproblem at iteration $m$ through ${\bfg {d}^{(m)}}$.

The objective function (\ref{mof1}) is identical to (\ref{eq1}) except for the last term, where ${c^{{\rm O},wc}}({\bfg v})$ is replaced with its approximation $\alpha$. Expression (\ref{mcompact}) includes the upper-level constraints. As per (\ref{mof3}), the operating cost corresponding to the uncertainty realizations identified at iteration $m$ represents a lower bound for $\alpha$. Constraints (\ref{mof4}) correspond to lower-level constraints (\ref{eq7}). Finally, the nonnegativity of $\alpha$ is imposed in (\ref{neg}).

\vspace{-0.5cm}
\subsection{Subproblem}\label{subproblem}
At each iteration $k$ of the column-and-constraint generation algorithm, the subproblem determines the worst-case uncertainty realizations yielding the maximum value of the minimum operating cost for a given upper-level decision provided by the previous master problem. Mathematically, the subproblem is a max-min problem comprising the second-stage problem (\ref{eq4})--(\ref{eq7}) parameterized in terms of the given upper-level decision variables ${\bfg {v}^{(k)}}$.

\reviewcol{Here, we propose solving such a bilevel problem through a decomposition-based method that was originally devised to deal with an instance of mathematical programming arising in structural reliability similar to (\ref{eqa1})--(\ref{eqa2}), as explained in \cite{MinguezCG:11}.} Thus, extending the application scope of this technique to handle a max-min problem constitutes a relevant methodological contribution of this work.

Based on \cite{MinguezCG:11}, the proposed method involves the iterative solution of two optimization problems: 1) the subproblem for middle-level variables, wherein lower-level variables $\bfg x$ are fixed in the max-min subproblem; and 2) the subproblem for lower-level variables, wherein middle-level variables $\bfg d$ are fixed in the max-min subproblem.

By fixing $\bfg d$ in the max-min subproblem at iteration $k$ of the column-and-constraint generation algorithm, the subproblem for lower-level variables at iteration $\nu$ of the inner loop is formulated as the following linear program:
\vspace{-0.2cm}
\begin{align}
&c^{{\rm O},min}({\coleq v}^{(k)},{\bfg d}^{(\nu)})=\underset{{\bfg d},{\bfg x}}{\text{Minimize}} \quad {{\bfg b}^T}{\bfg x}\label{eq111}\\
&\qquad\qquad\qquad\qquad\hspace{0.25cm}\text{subject to:}\notag\\
&\qquad\qquad\qquad\qquad\hspace{0.25cm}{\bfg x}\in \Omega({\bfg {v}^{(k)}},{\bfg d})\label{eq121}\\
&\qquad\qquad\qquad\qquad\hspace{0.25cm}{\bfg d}={\bfg {d}^{(\nu)}}: {\bfg \eta}\label{eq131}%\\
\end{align}
\noindent where the vector of dual variables $\bfg \eta$ represents the sensitivities of the optimal value of the objective function minimized in (\ref{eq111}), i.e., the minimum operating cost $c^{{\rm O},min}({\coleq v}^{(k)},{\bfg d}^{(\nu)})$, with respect to infinitesimal perturbations of the fixed vector of uncertainty realizations. As can be observed, problem (\ref{eq111})--(\ref{eq131}) comprises expressions (\ref{eq111}) and (\ref{eq121}), corresponding to the lower-level problem (\ref{eq6})--(\ref{eq7}) for given upper-level decisions ${\bfg {v}^{(k)}}$, as well as expression (\ref{eq131})  whereby middle-level variables ${\bfg d}$ are fixed.

Following the steps described in \cite{MinguezCG:11}, the fixed values of middle-level variables used in problem (\ref{eq111})--(\ref{eq131})  result from the optimal solution to the previous subproblem for middle-level variables. Note also that the application of the decomposition method presented in \cite{MinguezCG:11} relies on the availability of an analytical expression for the minimum operating cost, $c^{{\rm O},min}({\bfg v},{\bfg d})$, in terms of middle-level variables, ${\bfg d}$. In the absence of such an expression in the max-min subproblem, at each iteration of the inner loop, the subproblem for middle-level variables is built upon the first-order Taylor series approximation of the minimum operating cost around the uncertainty realizations identified at the previous iteration. Thus, at iteration $\nu$ of the inner loop, the subproblem for middle-level variables is cast as follows:
\begin{align}
&\underset{{\bfg d}}{\text{Maximize}} \quad {c^{{\rm O},min}({\coleq v}^{(k)},{\bfg d}^{(\nu-1)})} + {{{\bfg \eta}^{(\nu-1)}}^T}({\bfg d}-{\bfg {d}^{(\nu-1)}})   \label{of211}\\
&\text{subject to:}\notag\\
%&\text{Constraints (\ref{eq.inconsRCellip2})--(\ref{eq.inconsRCellip22})}.\label{compact21}%\\
&\left({\bfg d}-\bar{\bfg d}\right)^T \left({\bfg \Sigma}\right)^{-1}\left({\bfg d}-\bar{\bfg d}\right)\le \beta^2.\label{compact21}%\\
\end{align}

The approximate minimum operating cost maximized in (\ref{of211}) is based on the optimal minimum operating cost $c^{{\rm O},min}({\coleq v}^{(k)},{\bfg d}^{(\nu-1)})$ and the sensitivity vector ${\bfg \eta}^{(\nu-1)}$  previously obtained from (\ref{eq111})--(\ref{eq131}). The maximization of the approximate minimum operating cost is subject to middle-level constraint (\ref{eq5}), as formulated in (\ref{compact21}).

Based on \cite{Ackooij:17}, the analytical solution to problem (\ref{of211})--(\ref{compact21}) is given by:
\begin{align}
&{\bfg d}=\bar {\bfg d}+\beta\fraca{{\bfg \Sigma}{\bfg \eta}^{(\nu-1)}}{\sqrt{{{\bfg \eta}^{(\nu-1)}}^T{\bfg \Sigma}{\bfg \eta}^{(\nu-1)}}}\hspace{3pt}.\label{anali}%\\
\end{align}

Once initial values for middle-level variables ${\bfg d}$ are selected, the inner loop iteratively solves the linear program (\ref{eq111})--(\ref{eq131}) and the analytical expression (\ref{anali}) equivalent to (\ref{of211})--(\ref{compact21}), which are both computationally inexpensive. This iterative process terminates when the minimum operating cost remains unchanged within a user-defined tolerance.

\color{black}Due to the nonconvexity of the max-min subproblem, the inner loop only guarantees the attainment of a suboptimal solution, which  corresponds to a local optimum in case no degeneracy exists, i.e., if the gradient of the minimum operating cost is unique. Note that this type of local solution corresponds to the minimum of a linear cost function with one convex equality constraint, for which their gradients being parallel is the necessary and sufficient condition for local optimality. Nevertheless, if the iterative algorithm converges to a solution where degeneracy occurs, neither local nor global optimality can be ascertained but the algorithm might still provide appropriate solutions. In fact, the probability approximation (\ref{Probapprox}) still holds.
\color{black}
Thus, the proposed nested decomposition provides a lower bound for the optimal solution to the original trilevel problem. To verify the quality of the solution, different initial conditions have been tested for the inner loop. Our computational experience described in Section \ref{CaseStudy} revealed fast convergence and consistent results.
\vspace{-0.15cm}
\subsection{Algorithm}\label{Algorithm}
%\vspace{-0.15cm}

The proposed nested decomposition works as follows:

\begin{enumerate}
    \item \textit{Initialization of the outer loop.}
            \begin{itemize}
                \item Set the iteration counter of the outer loop $k$ to 1.
                \item Set the initial expansion plan ${\bfg {v}^{(k)}}={\bfg {0}}$.
            \end{itemize}
    \item \label{initial_descent} \textit{Initialization of the inner loop.} Set the iteration counter of the inner loop $\nu$ to 1 and select initial values for ${\bfg {d}^{(\nu)}}$.

    \item \label{sub_step1} \textit{Solution of problem (\ref{eq111})--(\ref{eq131}).} Solve problem (\ref{eq111})--(\ref{eq131}) for the given expansion plan ${\bfg {v}^{(k)}}$ and given ${\bfg {d}^{(\nu)}}$. This step provides ${\bfg {\eta}^{(\nu)}}$ and ${c^{{\rm O},min}}({\coleq v}^{(k)},{\bfg {d}^{(\nu)}})$. An approximation of the optimal value of the total cost minimized in (\ref{eq1}) can be computed as $C^{\rm SP}={{\bfg C^{\rm{L}}}}^T{\bfg {v}^{(k)}}+\sigma{c^{{\rm O},min}}({\coleq v}^{(k)},{\bfg {d}^{(\nu)}})$.

    \item \label{update_counters_inner} \textit{Update of the iteration counter of the inner loop.} Increase the iteration counter $\nu \leftarrow \nu+1$.

    \item \label{sub_step2} \textit{Solution of problem (\ref{of211})--(\ref{compact21}).} Solve problem (\ref{of211})--(\ref{compact21}) for given ${\bfg {d}^{(\nu-1)}}$ and ${\bfg {\eta}^{(\nu-1)}}$ through the equivalent analytical expression (\ref{anali}). This step provides ${\bfg {d}^{(\nu)}}$.

    \item \label{solution_check} \textit{Inner-loop convergence checking.} If uncertain parameters do not change sufficiently, i.e., $||{\bfg {d}^{(\nu)}}-{\bfg {d}^{(\nu-1)}}|| \leq \epsilon^{\rm {IL}}$, then set ${\bfg {d}^{(k)}}\leftarrow {\bfg {d}^{(\nu)}}$ and go to step \ref{update_counters}; otherwise, go to step \ref{sub_step1}.

    \item \label{update_counters} \textit{Update of the iteration counter of the outer loop.} Increase the iteration counter $k \leftarrow k+1$.

    \item \label{step_master} \textit{Master problem solution.} Solve the master problem (\ref{mof1})--(\ref{neg}) for given ${\bfg {d}^{(m)}}$, $m=1,\ldots, k-1$. This step provides ${\bfg {v}^{(k)}}$, $\alpha$, and an approximation of the optimal value of the total cost, namely $C^{\rm MP}={{{\bfg C^{\rm{L}}}}^T{\bfg {v}^{(k)}}+\sigma \alpha}$.

    \item \label{stop_step} \textit{Outer-loop convergence checking}. If a solution with a level of accuracy $\epsilon^{\rm {OL}}$ has been found, i.e., $(C^{\rm SP}-C^{\rm MP})/C^{\rm SP} \leq \epsilon^{\rm {OL}}$, the algorithm stops; otherwise, go to step \ref{initial_descent}.

\end{enumerate}

\section{Numerical Results}\label{CaseStudy}

Results from \color{black}three \color{black}case studies are presented in this section. An illustrative example built on that described in \cite{MinguezG:16} is first analyzed. The second case study is based on a modified version of the extended IEEE 118-bus system \cite{Pena:17}. \color{black}Finally, a benchmark based on the Polish 2383-bus system \cite{Polish} is examined. For the sake of reproducibility, input data for the case studies can be downloaded from \cite{data}. This repository also includes a figure depicting the 118-bus system. Both the illustrative and 118-bus test systems have been analyzed using synthetic data based on reality. Regarding the case study based on the Polish 2383-bus test system, we use wind data from 27 real locations in Spain where wind farms are installed, whereas demand correlation is synthetically generated using real correlation values from the Australian system \cite{BellWFH:15}.
In addition, numerical testing has been conducted for several values of the conservativeness parameter $\beta$.
%%%
%%% Numerical testing has been conducted for three correlation coefficients related to generating units,  \color{black}namely $0.5$, $0.7$, and $0.9$. In addition, three values of the conservativeness parameter $\beta$ have been considered, namely $0.84$, $1.28$, and $2.33$, which correspond to $\Phi(\beta)$ equal to $0.80$, $0.90$, and $0.99$, respectively.

%As mentioned in Section \ref{subproblem}, there is no guarantee that the inner loop identifies the actual worst-case operating cost for a given expansion plan. However, b
Based on the analogy presented in Section~\ref{s.analogy}, it is possible to gain confidence about the quality of the solution provided by the inner loop by an out-of-sample assessment relying on Monte Carlo simulation. To that end, for the expansion plan associated with each value of $\beta$, the lower-level problem (\ref{eq6})--(\ref{eq7}) has been solved for 10000 random samples of correlated uncertainty realizations {\bfg d}, thereby giving rise to a distribution function of sampled operating costs. The random values for {\bfg d} were generated using the following affine mapping {\cite{Ben-TalN:99}}:
\begin{equation}
  {\bfg d}=\bar {\bfg d}+ {\bfg L}{\bfg z}\label{eq.inconsRCellip2}
\end{equation}
\noindent where the mapping matrix ${\bfg L}$ was obtained from the Cholesky factorization of the variance-covariance matrix ${\bfg \Sigma}={\bfg L}{\bfg L}^{T}$, whereas the values of the perturbation vector ${\bfg z}$ were randomly sampled from an independent standard normal distribution. \color{black}
Thus, the quality of the subproblem solutions attained by the proposed approach is measured by the relative difference between $\Phi(\beta)$ and the probability that the sampled operating costs do not exceed the worst-case operating cost determined by the inner loop of the nested decomposition. \color{black}Such a probability is hereinafter referred to as sampled confidence level.
%Thus, the effectiveness of the proposed approach is measured by the relative difference between the $\Phi(\beta)$-quantile of the sampled operating costs and the worst-case operating cost determined by the inner loop of the nested decomposition.
%%%The probability of not exceeding the worst-case operating cost associated with the best expansion plan is equal to the percent number of successful samples, i.e., those featuring a minimum operating cost less than or equal to such a worst-case value. Thus, the effective performance of the proposed approach is backed when such a sampled probability is approximately equal to $\Phi(\beta)$.

For assessment purposes, we have also implemented the proposed approach while neglecting correlation in the optimization process. In other words, in spite of the existence of generation \color{black}and demand  correlations, their effect \color{black}is disregarded in the decision-making problem, as done in available models for robust transmission network expansion planning  \cite{WuCX:08}--\cite{ZhangC:17}. To that end, all off-diagonal entries of the variance-covariance matrix are set to $0$.

%%%We have implemented two models, the proposed approach and a similar one neglecting correlation. Model neglecting correlation is obtained considering the diagonal of the variance-covariance matrix $\Sigma$ in the proposed approach. We estimate the operating costs associated with the optimal expansion plan achieved neglecting correlation and consider correlation by means of a Monte Carlo sampling. Numerical testing has been conducted for different values of the conservativeness parameter $\beta$, namely 0.84, 1.28 and 2.33, which corresponds to confidence levels equal to 0.8, 0.9, and 0.99, respectively.

Simulations have been implemented on a Dell PowerEdge R920X64 with four Intel\textsuperscript{\textregistered} Xeon\textsuperscript{\textregistered} E7-4820 processors at 2.00 GHz and 768 GB of RAM using CPLEX 12.6 under GAMS 24.2 \cite{gams}. For all simulations, $\epsilon^{\rm {IL}}$, $\epsilon^{\rm {OL}}$, and the optimality tolerance for the branch-and-cut algorithm of CPLEX were set at 10$^{-8}$, \color{black}whereas, as done in \cite{Ruiz:15}, $\sigma$ was set equal to the number of hours in one year, i.e., 8760.\color{black}
%\vspace{-0.3cm}
\begin{figure}[t!]
  \begin{center}
  \includegraphics[width=0.34\textwidth]{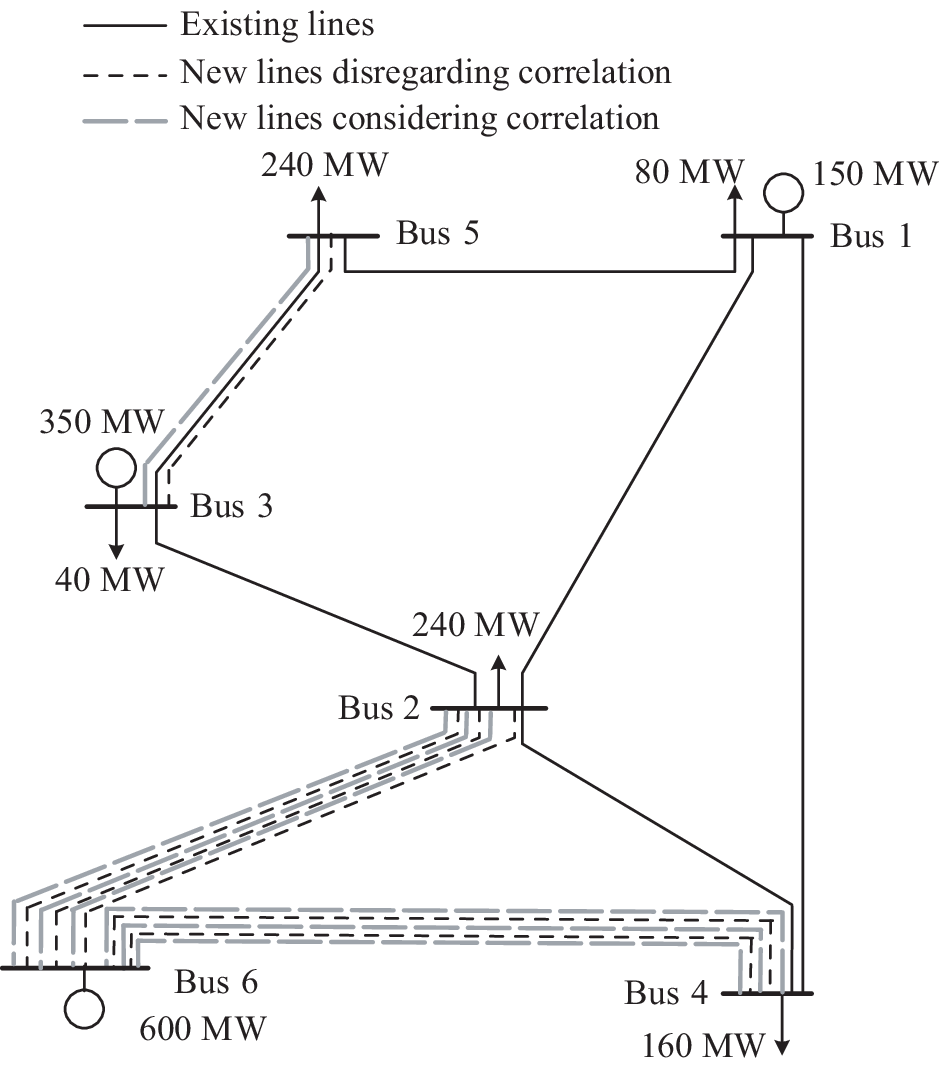}
    \caption{Illustrative example -- Expansion plans for $\beta=2.33$ and COR$\;=0.9$.}
    \label{6node}
  \end{center}
\vspace{-0.5cm}
\end{figure}

%
%\begin{figure}[t!]
%  \begin{flushleft}
%  \begin{minipage}{0.45\textwidth}
%  \includegraphics[width=\textwidth]{SimulacionGarver.eps}
%   \end{minipage}
%   \hspace*{-3.5cm}
%  \begin{minipage}{0.15\textwidth}
%  \vspace*{-0.5cm}
%  \includegraphics[width=\textwidth]{SimulacionGarverZOOM.eps}
% \end{minipage}
%    \caption{Illustrative example -- cumulative distribution functions of sampled operating costs (SOC) and points (SOC,$\Phi(\beta)$) for selected $\beta$-values associated with 0.9 correlation coefficient. Both implementations, considering correlation and disregarding correlation are considered.}
%    \label{6nodenormalprobplot}
%  \end{flushleft}
%\end{figure}
%
%\begin{figure}[t!]
%  \begin{flushleft}
%  \begin{minipage}{0.45\textwidth}
%  \includegraphics[width=\textwidth]{SimulacionGarver2.eps}
%   \end{minipage}
%   \hspace*{-3.5cm}
%  \begin{minipage}{0.15\textwidth}
%  \vspace*{-0.5cm}
%  \includegraphics[width=\textwidth]{SimulacionGarverZOOM2.eps}
% \end{minipage}
%    \caption{Illustrative example -- cumulative distribution functions of sampled operating costs (SOC) and points (SOC,$\Phi(\beta)$) for selected $\beta$-values associated with 0.7 correlation coefficient. Both implementations, considering correlation and disregarding correlation are considered.}
%    \label{6nodenormalprobplot2}
%  \end{flushleft}
%\end{figure}

\subsection{Illustrative Example}
The first test system comprises 6 buses, 3 generating units, 5 loads, and 6 existing lines (Fig. \ref{6node}). As described in \cite{data}, a \$4-million investment budget is imposed and each pair of buses can be connected by three lines at most, which amounts to 39 candidate lines. \color{black}All demands and generation capacities are sources of uncertainty. Fig.~\ref{6node} shows the mean or forecast values of demand and generation capacities. Demand correlation is accounted for as given in Table~\ref{garvercorr}, whereas a positive linear correlation between generating units located at buses 1 and 3 is considered. To that end,
three correlation coefficients, namely $0.5$, $0.7$, and $0.9$, have been examined. In addition, three values of the conservativeness parameter $\beta$ have been analyzed, namely $0.84$, $1.28$, and $2.33$, which correspond to $\Phi(\beta)$ equal to $0.80$, $0.90$, and $0.99$, respectively.\color{black}
\begin{table}[t!]
\caption{Illustrative Example -- Demand Correlation Matrix}
\centering
\color{black}
\begin{tabular}{ cccccc}
\hline\noalign{\smallskip}
Bus&1&2&3&4&5\\
\hline\noalign{\smallskip}
1&1.0 & 0.7&0.6 & 0.5&0.8\\
2&0.7 &1.0 & 0.8& 0.8&0.7\\
3& 0.6& 0.8& 1.0& 0.5&0.8\\
4&0.5 & 0.8& 0.5& 1.0&0.5\\
5& 0.8& 0.7& 0.8& 0.5&1.0\\
\hline\noalign{\smallskip}
\end{tabular}
\color{black}
\label{garvercorr}
\end{table}

%%%For expository purposes, correlation is only considered between pairs of generating units, for which a pre-specified correlation coefficient is used in the variance-covariance matrix.

%Demands are characterized by a $\pm20\%$ fluctuation divided by 2.32636 with respect to the corresponding nominal level, generation capacity for units located at buses 1 and 3 can oscillate by a maximum of 50\% divided by 2.32636 of their nominal values, while standard deviation of generation capacity for unit located at bus 6 has been set to 0.001.

For all instances, convergence was attained after 5 outer-loop iterations at most in less than 5 s. The results are summarized in Tables \ref{garver} and \ref{garveroos}, where COR stands for generation correlation coefficient, IC and SOC respectively denote the investment cost, ${{\bfg C^{\rm{L}}}}^T{\bfg v}$, and the scaled worst-case operating cost, $\sigma c^{{\rm O},wc}({\bfg v})$, provided by the nested decomposition, and SCL represents the sampled confidence level for the worst-case operating cost.

%\textbf{increases as the confidence level increases} (\textbf{ESTO NO SE CORRESPONDE CON LA TABLA}), \textbf{achieving a maximum reduction of}

%and $\Phi(\beta)$-quantile represents the sampled operating cost associated with probability $\Phi(\beta)$.
%In addition, Fig. \ref{6nodenormalprobplot} shows cumulative distribution functions of sampled operating costs (SOC) and points (($SOC,\Phi(\beta)$), which are plotted using different markers: \color{grismedio}$\rhd$\color{black}, \color{grismedio}$\circ$\color{black}, \color{grisclaro}$\ast$\color{black}, \color{grisclaro}$\diamond$\color{black}, $\square$, $\times$, also shown in Table \ref{garver}), for selected $\beta$-values associated with 0.9 correlation coefficient. Note that we are using a normal probability plot.

\begin{table}[t!]
\caption{Illustrative Example -- Investment Costs, Scaled Worst-Case Operating Costs, and Sampled Confidence Levels}
\centering
\color{black}
\renewcommand{\tabcolsep}{5pt}
\begin{tabular}{cccccccc}
\cline{3-8}\multirow{1}{*}{}&\multirow{1}{*}{}&\multicolumn{3}{c}{Disregarding correlation}&\multicolumn{3}{c}{Considering correlation}\\
\hline
$\beta$&\multirow{2}{*}{COR}& IC  & SOC  &\multirow{2}{*}{SCL}& IC  & SOC  &\multirow{2}{*}{SCL}\\
($\Phi(\beta)$)&&(10$^6$ \$)&(10$^6$ \$)& &(10$^6$ \$)&(10$^6$ \$)&\\
\hline
\multirow{2}{*}{0.84}&0.5&2.70&333.73&0.65& 3.09& \hspace{1.4mm}345.14&0.79\\
\multirow{2}{*}{(0.80)}&0.7&2.70&333.73&0.65&3.09 &\hspace{1.4mm}346.08 &0.80\\
&0.9&2.70&333.33&0.65& 3.09&\hspace{1.4mm}346.99 &0.81\\
\hline
\multirow{2}{*}{1.28}&0.5&2.70&348.14&0.77& 3.67& \hspace{1.4mm}365.51&0.90\\
\multirow{2}{*}{(0.90)}&0.7&2.70&348.14&0.77&3.67 &\hspace{1.4mm}366.94 &0.91\\
&0.9&2.70&348.14&0.77& 3.67&\hspace{1.4mm}368.33&0.96\\
\hline
\multirow{2}{*}{2.33}&0.5&3.28&382.37&0.96& 3.86& 1206.55&0.99\\
\multirow{2}{*}{(0.99)}&0.7&3.28&382.37&0.96&3.86 &\hspace{1.4mm}960.97 &0.99\\
&0.9&3.28&382.37&0.96& 3.86&1781.68 &0.99\\
\hline
\end{tabular}
\color{black}
\label{garver}
\end{table}

\begin{table}[t!]
    \centering
    \caption{Illustrative example -- Quantiles of Scaled Operating Costs Resulting from the Out-of-Sample Assessment (10$^6$ \$) }
    \color{black}
         \renewcommand{\tabcolsep}{1pt}
     \hspace*{-0.2cm}
\begin{tabular}{
%@{\extracolsep{2pt}}c@{\extracolsep{2pt}}c@{\extracolsep{3pt}}c@{\extracolsep{3pt}}l@{\extracolsep{3pt}}c@{\extracolsep{3pt}}c@{\extracolsep{3pt}}c@{\extracolsep{3pt}}c@{\extracolsep{3pt}}c@{\extracolsep{3pt}}c
cccccccccc
}
\cline{3-10}\noalign{\smallskip}\multirow{1}{*}{}&\multirow{1}{*}{}&\multicolumn{4}{c}{Disregarding correlation}&\multicolumn{4}{c}{Considering correlation}\\
\cline{3-10}\noalign{\smallskip}\multirow{1}{*}{}&\multirow{1}{*}{}&\multicolumn{4}{c}{Confidence level}&\multicolumn{4}{c}{Confidence level}\\
\hline\noalign{\smallskip}
$\beta$ & \multirow{2}{*}{COR} & \multirow{2}{*}{0.50}  & \multirow{2}{*}{0.80} & \multirow{2}{*}{0.90} & \multirow{2}{*}{0.99} &  \multirow{2}{*}{0.50}  & \multirow{2}{*}{0.80} & \multirow{2}{*}{0.90} & \multirow{2}{*}{0.99}\\
($\Phi(\beta)$)&& & & & & & & & \\
\hline\noalign{\smallskip}
\multirow{2}{*}{0.84}&0.5&321.66&359.21&1974.15&8242.63& 320.24& 345.63& 1337.56&7933.45 \\
\multirow{2}{*}{(0.80)}&0.7&321.63&358.31&1960.74&8252.21&320.18 &345.58 & 1293.96& 7952.65\\
&0.9&321.76&358.03&1990.51&8655.98& 320.22& 345.60& 1340.14&8306.57\\
\hline
\multirow{2}{*}{1.28}&0.5&321.66&359.21&1974.15&8242.63& 312.67& 341.51& \hspace{1.4mm}363.28&5959.08 \\
\multirow{2}{*}{(0.90)}&0.7&321.63&358.31&1960.74&8252.21&312.57 &341.70 & \hspace{1.4mm}362.57& 5821.00\\
&0.9&321.76&358.03&1990.51&8655.98& 312.82& 335.22& \hspace{1.4mm}348.29&\hspace{1.4mm}386.31\\
\hline
\multirow{2}{*}{2.33}&0.5&322.66&349.76&\hspace{1.4mm}365.10&1413.26& 321.64& 348.13& \hspace{1.4mm}363.27&\hspace{1.4mm}404.26 \\
\multirow{2}{*}{(0.99)}&0.7&322.50&349.91&\hspace{1.4mm}365.28&1491.43&321.61 &348.14 & \hspace{1.4mm}363.20& \hspace{1.4mm}407.62\\
&0.9&322.65&350.12&\hspace{1.4mm}365.42&1536.34& 321.75& 348.42& \hspace{1.4mm}363.24&\hspace{1.4mm}407.77\\
\hline\noalign{\smallskip}
\end{tabular}
\color{black}
\label{garveroos}
\end{table}

From Table \ref{garver}, the main observations are as follows:
\begin{enumerate}

%%%\item \label{obs1} For the lowest value of the conservativeness parameter, $\beta=0.84$, correlation has no impact on expansion planning. Thus, investment decisions are all identical, thereby leading to the same investment cost. Note that, when correlation is disregarded, additional investments are only required for the most conservative instance, i.e., for $\beta=2.33$. In contrast, when correlation is considered, as $\beta$ grows so does the investment cost.
    %Note that SOC cumulative distribution function curves shown in Fig \ref{6nodenormalprobplot} associated with $\beta=0.84$ (continuous and dashed medium gray lines) are almost coincident. In this particular case, expansion plan for $\beta=1.28$ without considering correlation also coincide. This is the reason why its SOC cumulative distribution curve (continuous light gray line) shown in Fig \ref{6nodenormalprobplot} is almost coincident with respect to those for $\beta=0.84$.

%\item All SOC cumulative distribution function curves shown in Fig \ref{6nodenormalprobplot} present an abrupt change of slope at different thresholds between \$200 million and \$400 million. We have tested numerically that this change is due to the need for load shedding.
    %, i.e. points on the left side of the threshold correspond to sampled operating costs without load shedding, while points on the right side of the threshold correspond to sampled operating costs with load shedding.

\item Investment costs and worst-case operating costs incurred when correlation is properly considered are respectively greater than or equal to those obtained when such a modeling aspect is ignored. \color{black}This result may misleadingly suggest that the investment plans obtained disregarding correlation are convenient. However, such solutions are suboptimal, as indicated by SCL being significantly lower than the corresponding expected confidence levels given by $\Phi(\beta)$.\color{black}

\item When correlation is considered, small differences are observed between the sampled confidence levels and the corresponding values of $\Phi(\beta)$, thereby revealing the high accuracy of the resulting worst-case operating costs.
    %Note that such differences decrease as $\beta$ grows, reaching negligible values for the most conservative instance. %For instance, for the case $\beta=2.33$, this difference is represented by the horizontal distance between the black $\times$ dot and the black dashed-line in Fig. \ref{6nodenormalprobplot}.
    \color{black}Note that, in general, such differences decrease as $\beta$ grows, reaching negligible values for the most conservative instance.  \color{black}Moreover, such errors are significantly lower than those incurred when correlation is disregarded. These results evidence the need for properly modeling correlation in case it exists as well as the suitability of the proposed approach to address the max-min subproblem.

\item As expected, the higher the conservativeness parameter $\beta$ is, the higher the total cost is.
%%%\item For each value of $\beta$, when correlation is modeled, the $\Phi(\beta)$-quantile of the sampled operating costs grows as the correlation coefficient increases. In contrast, when correlation is ignored, no pattern is followed.

\end{enumerate}

 As illustrated in Table \ref{garveroos}, the out-of-sample assessment based on Monte Carlo simulation is also useful to show the advantages of considering correlated uncertainty over disregarding correlation, as done in existing robust expansion planning models \cite{WuCX:08}--\cite{ZhangC:17}. For the investment solutions to all instances, Table~\ref{garveroos} lists the scaled values of the sampled operating costs for confidence levels equal to 0.50, 0.80, 0.90, and 0.99 and the same conservativeness parameters and correlation coefficients analyzed in Table \ref{garver}.
%%%The identical expansion plans identified for $\beta=0.84$, both disregarding and considering correlation, and for $\beta=1.28$ and neglecting correlation featured the same cost quantiles for each confidence level and correlation coefficient.
\color{black}As can be observed, accounting for correlation gives rise to lower $\Phi(\beta)$-quantiles. Note that reduction factors of up to 95.5\% are reached, as is the case for $\beta=1.28$ and a confidence level equal to 0.99. Overall, considering correlation is economically appealing as it yields the lowest values for the sum of IC and the $\Phi(\beta)$-quantile. These results show the superiority of the proposed approach over existing robust models neglecting correlation, thereby corroborating the relevance of considering such a practical modeling feature\color{black}.

In order to show the impact of correlation on expansion planning, a representative instance is selected corresponding to $\beta$ equal to $2.33$ and a correlation coefficient equal to $0.9$. For such an instance, Fig. \ref{6node} depicts the expansion plans identified by the proposed approach when correlation is disregarded in the optimization process and when this modeling aspect is properly accounted for. Neglecting correlation leads to the construction of three lines in corridor 2-6, one line in corridor 3-5, and two lines in corridor 4-6. The investment cost is equal to \$3.28 million whereas the $\Phi(\beta)$-quantile of the scaled operating costs resulting from the out-of-sample assessment amounts to \color{black}\$1536.34 \color{black}million. Such a sampled estimation of the scaled worst-case operating cost is 4 times larger than the value provided by the optimization, i.e., \$382.37 million, due to the need for load shedding. In contrast, accounting for correlation requires the construction of an additional line in corridor 4-6 (Fig. \ref{6node}). Note that the resulting 17.7\% increase in investment cost is clearly offset by the substantial \color{black}73.5\% \color{black}reduction in the $\Phi(\beta)$-quantile, which is as low as \color{black}\$407.77 \color{black}million because load shedding is prevented.

\subsection{IEEE 118-Bus System}
The second case study comprises 118 buses, \color{black}40 \color{black}conventional generators, \color{black}40 \color{black}wind farms, 91 loads, and 180 existing transmission lines. A set of 72 candidate lines is available under a \$20-million investment budget. All demand consumptions and all generation capacities of wind farms are uncertain. System data can be found in \cite{data}. A positive linear correlation is considered between four pairs of generators, namely 11-12, 36-78, 40-41, and 79-80. \color{black}Additionally, the demands of consecutively numbered buses are positively correlated. For the sake of simplicity, a correlation coefficient equal to 0.5 is used for demands.\color{black}
\begin{table}[t!]
    \centering
    \caption{IEEE 118-Bus System -- Investment Costs, Scaled Worst-Case Operating Costs, and Sampled Confidence Levels}
\color{black}
\renewcommand{\tabcolsep}{5pt}
\begin{tabular}{cccccccc}
\cline{3-8}\multirow{1}{*}{}&\multirow{1}{*}{}&\multicolumn{3}{c}{Disregarding correlation}&\multicolumn{3}{c}{Considering correlation}\\
\hline
$\beta$&\multirow{2}{*}{COR}& IC  & SOC  &\multirow{2}{*}{SCL}& IC  & SOC  &\multirow{2}{*}{SCL}\\
($\Phi(\beta)$)&&(10$^6$ \$)&(10$^6$ \$)& &(10$^6$ \$)&(10$^6$ \$)&\\
\hline
\multirow{2}{*}{0.84}&0.5&5.80&2116.21&0.72& 3.96& 2078.04&0.75\\
\multirow{2}{*}{(0.80)}&0.7&5.80&2116.21&0.72 & 3.61 &2079.01 &0.75\\
&0.9&5.80&2116.21&0.72& 3.61&2079.40 &0.75\\
\hline
\multirow{2}{*}{1.28}&0.5&5.15&2145.45&0.81 & 5.55& 2102.21&0.88\\
\multirow{2}{*}{(0.90)}&0.7&5.15&2145.45&0.81&5.56 &2102.33 &0.90\\
&0.9&5.15&2145.45&0.81& 5.65&2102.60&0.88\\
\hline
\multirow{2}{*}{2.33}&0.5&7.71&2213.40&0.97& 9.93& 2166.77&0.99\\
\multirow{2}{*}{(0.99)}&0.7&7.71&2213.40&0.97&9.93 &2168.40 &0.99\\
&0.9&7.75&2213.40&0.97&9.93&2170.01 &0.99\\
\hline
\end{tabular}
\color{black}
\label{118}
\end{table}

\begin{table}[t!]
    \centering
    \caption{IEEE 118-Bus System -- Quantiles of Scaled Operating Costs Resulting from the Out-of-Sample Assessment (10$^6$ \$) }
    \color{black}
     \renewcommand{\tabcolsep}{1pt}
     \hspace*{-0.6cm}
\begin{tabular}{
cccccccccc
}
\cline{3-10}\noalign{\smallskip}\multirow{1}{*}{}&\multirow{1}{*}{}&\multicolumn{4}{c}{Disregarding correlation}&\multicolumn{4}{c}{Considering correlation}\\
\cline{3-10}\noalign{\smallskip}\multirow{1}{*}{}&\multirow{1}{*}{}&\multicolumn{4}{c}{Confidence level}&\multicolumn{4}{c}{Confidence level}\\
\hline\noalign{\smallskip}
$\beta$ & \multirow{2}{*}{COR} & \multirow{2}{*}{0.50}  & \multirow{2}{*}{0.80} & \multirow{2}{*}{0.90} & \multirow{2}{*}{0.99} &\multirow{2}{*}{0.50}  & \multirow{2}{*}{0.80} & \multirow{2}{*}{0.90} & \multirow{2}{*}{0.99}\\
($\Phi(\beta)$)&& & & & & & & & \\
\hline\noalign{\smallskip}
\multirow{2}{*}{0.84}&0.5&2070.94&2142.99&2719.57&10898.60& 2038.18& 2094.41& 3335.34&11008.10 \\
\multirow{2}{*}{(0.80)}&0.7&2071.14&2143.34&2777.76&11145.90&2038.40 &2094.32& 3334.19& 10730.30\\
&0.9&2070.90&2142.60&2686.65&10820.50& 2038.19& 2093.91& 3301.41&10655.10\\
\hline
\multirow{2}{*}{1.28}&0.5&2071.54&2141.18&2504.87&10580.60&2035.31& 2077.96& 2117.29&\hspace{1.4mm}7778.03 \\
\multirow{2}{*}{(0.90)}&0.7&2071.64&2141.65&2557.05&10781.70&2034.94 &2075.84 & 2104.51& \hspace{1.4mm}7783.56\\
&0.9&2071.44&2140.85&2473.01&10467.80& 2035.41& 2078.64& 2119.93&\hspace{1.4mm}7982.29\\
\hline
\multirow{2}{*}{2.33}&0.5&2067.96&2123.67&2154.46&\hspace{1.4mm}3617.70& 2034.42& 2071.80& 2092.14&\hspace{1.4mm}2168.53 \\
\multirow{2}{*}{(0.99)}&0.7&2068.07&2124.06&2154.94&\hspace{1.4mm}3795.07&2034.42 &2071.87 & 2092.33& \hspace{1.4mm}2171.60\\
&0.9&2067.98&2123.54&2154.17&\hspace{1.4mm}3484.76&2034.31& 2071.72& 2092.09&\hspace{1.4mm}2167.83\\
\hline\noalign{\smallskip}
\end{tabular}
\vspace*{-0.4cm}
\color{black}
\label{118oos}
\end{table}

Tables \ref{118} and \ref{118oos} summarize the results attained for the same conservativeness parameters and generation correlation coefficients analyzed in the illustrative example. For all instances, convergence was achieved after 7 outer-loop iterations at most in less than 4 hours. \color{black} From Tables \ref{118} and \ref{118oos}, the main observations are as follows:
\begin{enumerate}
\item 	For $\beta=0.84$, both investment and worst-case operating costs incurred when correlation is properly considered are lower than those obtained when such a modeling aspect is ignored. This result reveals the beneficial impact of accounting for correlation. According to the results given in Table~\ref{118oos}, the solution considering correlation is indeed advantageous over that disregarding correlation for quantiles less than or equal to the selected confidence level $\Phi(\beta)$. For quantiles of higher confidence levels, larger operating costs may be attained as the investment reduction comes at the expense of increasing the operating cost in the upper tail of its distribution. Note that this result is consistent with the choice for $\beta$.

\item 	For $\beta=1.28$ and $\beta=2.33$, the consideration of correlation gives rise to higher investment costs. Notwithstanding, such a cost increase is offset by the reduction in worst-case operating costs. The results shown in Table~\ref{118oos} confirm that expansion plans considering correlation are economically more convenient because all cost quantiles are lower than those obtained while disregarding correlation.

\item 	As compared with the model disregarding correlation, considering this practical aspect yields values of SCL closer to the desired $\Phi(\beta)$ and hence more accurate worst-case operating costs. Moreover, as $\beta$ increases, the accuracy of the worst-case operating costs improves as the gap between SCL and the corresponding $\Phi(\beta)$ is reduced.

\item As expected, as the conservativeness parameter $\beta$ grows, so does the total cost.

\end{enumerate}

As an example, \color{black}for $\beta=2.33$  and a generation correlation coefficient equal to $0.9$, the 0.99-quantile of the sampled operating costs for the expansion plan identified when correlation is considered is 1.6 times lower than that achieved when correlation is disregarded (Table \ref{118oos}). This is so because the \$9.93 million investment resulting from considering correlation is sufficient to prevent the need for load shedding under correlated uncertainty. Unfortunately, disregarding correlation gives rise to optimistic solutions which assume that no load shedding occurs when indeed it does. As shown in Table \ref{118oos}, the $\Phi(\beta)$-quantile of the scaled operating costs resulting from the out-of-sample assessment for the expansion plan identified while disregarding correlation, \$3484.76 million, represents a \$1316.93 million increase over that associated with the solution considering correlated uncertainty, \$2167.83 million. Such an increase in the $\Phi(\beta)$-quantile substantially exceeds the \$2.18 million reduction in the investment cost (Table \ref{118}), thus backing the need for properly accounting for correlation.

%the solution considering correlation gave rise to load shedding (as numerically checked). Note that in that particular case the level of investment does not allow avoiding load shedding. Notwithstanding, observations \ref{obs1}--\ref{obs4} described for the illustrative example also hold for this case study.

%An out-of-sample assessment is also presented for this case study to back the superiority of the proposed approach over previously reported expansion planning models disregarding correlation. Results are shown in Table \ref{118oos} and the same remarks presented in the illustrative example are also valid for this case study.
%It is worth mentioning that, unlike in the previous benchmark, neglecting correlation did not give rise to load shedding.

%
\begin{table}[htb]
\renewcommand{\arraystretch}{1.0}
\renewcommand{\tabcolsep}{2.0pt}
\begin{center}
\caption{2383-Bus System -- Correlations of Maximum Wind Power Productions}\label{WindfarmPW}
\begin{scriptsize}
\color{black}
\rotatebox{90}{
\begin{tabular}{cR{0.55cm}R{0.55cm}R{0.55cm}R{0.55cm}R{0.55cm}R{0.55cm}R{0.55cm}R{0.55cm}R{0.55cm}R{0.55cm}R{0.55cm}R{0.55cm}R{0.55cm}R{0.55cm}R{0.55cm}R{0.55cm}R{0.55cm}R{0.55cm}R{0.55cm}R{0.55cm}R{0.55cm}R{0.55cm}R{0.55cm}R{0.55cm}R{0.55cm}R{0.55cm}R{0.55cm}R{0.55cm}}
\hline
COR&\multicolumn{1}{c}{1}&\multicolumn{1}{c}{2}&\multicolumn{1}{c}{3}&\multicolumn{1}{c}{4}&\multicolumn{1}{c}{5}&\multicolumn{1}{c}{6}&
\multicolumn{1}{c}{7}&\multicolumn{1}{c}{8}&\multicolumn{1}{c}{9}&\multicolumn{1}{c}{10}&\multicolumn{1}{c}{11}&\multicolumn{1}{c}{12}&
\multicolumn{1}{c}{13}&\multicolumn{1}{c}{14}&\multicolumn{1}{c}{15}&\multicolumn{1}{c}{16}&\multicolumn{1}{c}{17}&\multicolumn{1}{c}{18}&
\multicolumn{1}{c}{19}&\multicolumn{1}{c}{20}&\multicolumn{1}{c}{21}&\multicolumn{1}{c}{22}&\multicolumn{1}{c}{23}&\multicolumn{1}{c}{24}&
\multicolumn{1}{c}{25}&\multicolumn{1}{c}{26}&\multicolumn{1}{c}{27}\\\hline
1$\;$&{\bf 1.00}&0.27&0.06&0.88&0.19&0.39&0.66&0.76&0.63&0.50&0.44&0.45&0.48&0.49&0.34&0.28&0.33&0.21&0.19&0.27&0.76&0.71&0.56&0.81&0.73&0.37&0.75\\
2$\;$&0.27&{\bf 1.00}&0.34&0.36&0.11&0.12&0.44&0.27&0.32&0.25&0.25&0.27&0.26&0.25&0.11&-0.01&0.54&0.12&0.12&0.27&0.25&0.23&0.28&0.16&0.23&0.16&0.20\\
3$\;$&0.06&0.34&{\bf 1.00}&0.12&0.01&-0.07&0.20&0.02&0.12&-0.02&-0.02&0.04&0.01&0.00&-0.02&-0.10&0.23&-0.04&-0.01&0.06&0.06&-0.00&0.09&0.01&-0.01&0.03&0.00\\
4$\;$&0.88&0.36&0.12&{\bf 1.00}&0.20&0.35&0.70&0.73&0.65&0.53&0.45&0.47&0.51&0.49&0.34&0.24&0.42&0.21&0.19&0.32&0.73&0.65&0.53&0.72&0.67&0.37&0.70\\
5$\;$&0.19&0.11&0.01&0.20&{\bf 1.00}&-0.05&0.17&0.23&0.27&0.23&0.24&0.25&0.22&0.26&0.33&0.51&0.24&-0.01&0.02&0.65&0.28&0.23&0.23&0.23&0.21&0.73&0.17\\
6$\;$&0.39&0.12&-0.07&0.35&-0.05&{\bf 1.00}&0.34&0.41&0.31&0.52&0.57&0.53&0.51&0.53&0.17&0.11&0.18&0.69&0.66&0.18&0.30&0.34&0.13&0.39&0.39&0.14&0.49\\
7$\;$&0.66&0.44&0.20&0.70&0.17&0.34&{\bf 1.00}&0.74&0.83&0.59&0.47&0.58&0.60&0.50&0.34&0.09&0.56&0.26&0.26&0.33&0.62&0.61&0.50&0.59&0.65&0.32&0.72\\
8$\;$&0.76&0.27&0.02&0.73&0.23&0.41&0.74&{\bf 1.00}&0.80&0.60&0.50&0.52&0.57&0.53&0.43&0.32&0.39&0.25&0.24&0.34&0.75&0.85&0.56&0.78&0.88&0.43&0.91\\
9$\;$&0.63&0.32&0.12&0.65&0.27&0.31&0.83&0.80&{\bf 1.00}&0.56&0.46&0.55&0.55&0.50&0.44&0.23&0.47&0.21&0.21&0.35&0.67&0.67&0.54&0.66&0.69&0.43&0.75\\
10$\;\;$&0.50&0.25&-0.02&0.53&0.23&0.52&0.59&0.60&0.56&{\bf 1.00}&0.65&0.71&0.96&0.60&0.36&0.15&0.36&0.49&0.43&0.48&0.50&0.46&0.29&0.50&0.52&0.40&0.63\\
11$\;\;$&0.44&0.25&-0.02&0.45&0.24&0.57&0.47&0.50&0.46&0.65&{\bf 1.00}&0.74&0.65&0.84&0.27&0.22&0.44&0.60&0.57&0.44&0.40&0.43&0.26&0.42&0.46&0.43&0.51\\
12$\;\;$&0.45&0.27&0.04&0.47&0.25&0.53&0.58&0.52&0.55&0.71&0.74&{\bf 1.00}&0.75&0.72&0.29&0.10&0.46&0.57&0.55&0.47&0.43&0.41&0.25&0.44&0.46&0.42&0.56\\
13$\;\;$&0.48&0.26&0.01&0.51&0.22&0.51&0.60&0.57&0.55&0.96&0.65&0.75&{\bf 1.00}&0.60&0.34&0.11&0.38&0.50&0.44&0.47&0.49&0.43&0.28&0.48&0.48&0.39&0.61\\
14$\;\;$&0.49&0.25&0.00&0.49&0.26&0.53&0.50&0.53&0.50&0.60&0.84&0.72&0.60&{\bf 1.00}&0.29&0.25&0.48&0.49&0.50&0.41&0.44&0.46&0.31&0.46&0.49&0.44&0.54\\
15$\;\;$&0.34&0.11&-0.02&0.34&0.33&0.17&0.34&0.43&0.44&0.36&0.27&0.29&0.34&0.29&{\bf 1.00}&0.40&0.15&0.05&0.03&0.35&0.43&0.35&0.26&0.42&0.36&0.43&0.42\\
16$\;\;$&0.28&-0.01&-0.10&0.24&0.51&0.11&0.09&0.32&0.23&0.15&0.22&0.10&0.11&0.25&0.40&{\bf 1.00}&0.14&-0.03&-0.01&0.26&0.34&0.39&0.28&0.36&0.35&0.52&0.29\\
17$\;\;$&0.33&0.54&0.23&0.42&0.24&0.18&0.56&0.39&0.47&0.36&0.44&0.46&0.38&0.48&0.15&0.14&{\bf 1.00}&0.19&0.22&0.33&0.32&0.36&0.35&0.23&0.36&0.29&0.33\\
18$\;\;$&0.21&0.12&-0.04&0.21&-0.01&0.69&0.26&0.25&0.21&0.49&0.60&0.57&0.50&0.49&0.05&-0.03&0.19&{\bf 1.00}&0.93&0.22&0.16&0.18&0.05&0.20&0.23&0.09&0.31\\
19$\;\;$&0.19&0.12&-0.01&0.19&0.02&0.66&0.26&0.24&0.21&0.43&0.57&0.55&0.44&0.50&0.03&-0.01&0.22&0.93&{\bf 1.00}&0.23&0.14&0.18&0.05&0.17&0.22&0.12&0.28\\
20$\;\;$&0.27&0.27&0.06&0.32&0.65&0.18&0.33&0.34&0.35&0.48&0.44&0.47&0.47&0.41&0.35&0.26&0.33&0.22&0.23&{\bf 1.00}&0.33&0.28&0.25&0.27&0.28&0.70&0.29\\
21$\;\;$&0.76&0.25&0.06&0.73&0.28&0.30&0.62&0.75&0.67&0.50&0.40&0.43&0.49&0.44&0.43&0.34&0.32&0.16&0.14&0.33&{\bf 1.00}&0.71&0.68&0.90&0.72&0.45&0.73\\
22$\;\;$&0.71&0.23&-0.00&0.65&0.23&0.34&0.61&0.85&0.67&0.46&0.43&0.41&0.43&0.46&0.35&0.39&0.36&0.18&0.18&0.28&0.71&{\bf 1.00}&0.62&0.73&0.96&0.41&0.79\\
23$\;\;$&0.56&0.28&0.09&0.53&0.23&0.13&0.50&0.56&0.54&0.29&0.26&0.25&0.28&0.31&0.26&0.28&0.35&0.05&0.05&0.25&0.68&0.62&{\bf 1.00}&0.56&0.60&0.35&0.50\\
24$\;\;$&0.81&0.16&0.01&0.72&0.23&0.39&0.59&0.78&0.66&0.50&0.42&0.44&0.48&0.46&0.42&0.36&0.23&0.20&0.17&0.27&0.90&0.73&0.56&{\bf 1.00}&0.75&0.43&0.79\\
25$\;\;$&0.73&0.23&-0.01&0.67&0.21&0.39&0.65&0.88&0.69&0.52&0.46&0.46&0.48&0.49&0.36&0.35&0.36&0.23&0.22&0.28&0.72&0.96&0.60&0.75&{\bf 1.00}&0.40&0.86\\
26$\;\;$&0.37&0.16&0.03&0.37&0.73&0.14&0.32&0.43&0.43&0.40&0.43&0.42&0.39&0.44&0.43&0.52&0.29&0.09&0.12&0.70&0.45&0.41&0.35&0.43&0.40&{\bf 1.00}&0.39\\
27$\;\;$&0.75&0.20&0.00&0.70&0.17&0.49&0.72&0.91&0.75&0.63&0.51&0.56&0.61&0.54&0.42&0.29&0.33&0.31&0.28&0.29&0.73&0.79&0.50&0.79&0.86&0.39&{\bf 1.00}\\\hline
\end{tabular}
}\color{black}
\vspace{-0.3cm}
\end{scriptsize}
\end{center}
\end{table}

\color{black}
\subsection{Polish 2383-Bus Test System}

This test system comprises 2383 buses, 327 generating units, 1500 loads, and 2896 existing lines. Expansion decisions are made on a set of 124 prospective lines under a \$20-million investment budget.
All demand consumptions and all generation capacities of wind farms (101 out of 327 generating units) are uncertain.

\begin{table}[t!]
    \centering
    \caption{2383-Bus System -- Investment Costs, Scaled Worst-Case Operating Costs, and Sampled Confidence Levels}
\color{black}
\begin{tabular}{ccccccc}
\cline{2-7}\noalign{\smallskip}\multirow{1}{*}{}&\multicolumn{3}{c}{Disregarding correlation}&\multicolumn{3}{c}{Considering correlation}\\
\hline\noalign{\smallskip}
\multirow{2}{*}{$\beta$}& IC  & SOC  &\multirow{2}{*}{SCL}& IC  & SOC  &\multirow{2}{*}{SCL}\\
&(10$^6$ \$)&(10$^6$ \$)& &(10$^6$ \$)&(10$^6$ \$)&\\
\hline\noalign{\smallskip}
\multirow{1}{*}{2.32}&4.41&4915.53&0.44& 10.78& 12440.54 &0.99\\
\hline\noalign{\smallskip}
\end{tabular}
\color{black}
\label{2383}
\end{table}

All demands are positively correlated with a correlation coefficient equal to 0.7.
Regarding correlated wind uncertainty, for a subset of 27 wind farms, we have considered the correlations in 27 real locations in Spain. Wind data are obtained from the SeaWind database \cite{MenendezGFFMG:13}, which constitutes an hourly wind reanalysis over a 15 km spatial resolution grid for the entire 1989--2009
period, covering the South Atlantic European region and the Mediterranean basin. We have transformed wind speed into power production by using the power curve of an NREL 5-MW turbine \cite{JonkmanBMS:09}. Correlations among maximum wind farm power productions are given in Table~\ref{WindfarmPW}. As can be observed, the maximum wind farm productions are mostly positively correlated, being the magnitude of such correlations related to the geographical distance, location, and climate conditions. Note that 75\% of the correlation values are above 0.23, 50\% are above 0.39, 25\% are above 0.54, and 5\% are above 0.76, reaching values of up to 0.96 for close geographical locations (within 100 km).  The interested reader is referred to \cite{data} for a full description of the benchmark.

For the sake of conciseness, we report results from a single value of the conservativeness parameter $\beta$, namely $2.32$, which corresponds to $\Phi(\beta)$ equal to $0.99$. Tables~\ref{2383} and \ref{2383oos} summarize the results. The nested decomposition converged in 59.5 hours for the instance considering correlation.

As can be observed in Table~\ref{2383}, considering correlation leads to higher values of IC and SOC while featuring a value of SCL substantially closer to the desired confidence level associated with the selected conservativeness parameter. The economic advantage of such a greater level of accuracy is verified by the quantiles of scaled operating costs shown in Table~\ref{2383oos}. Note that, for all confidence levels, the expansion plan considering correlation features lower cost quantiles. Moreover, such cost reductions, which are above \$89.11 million, compensate for the \$6.37 million increase in investment cost. Thus, the expansion plan considering correlation outperforms the investment scheme neglecting this aspect.

\begin{table}[t!]
    \centering
    \caption{2383-Bus System -- Quantiles of Scaled Operating Costs Resulting from the Out-of-Sample Assessment (10$^6$ \$) }
    \color{black}
     \renewcommand{\tabcolsep}{1pt}
%     \hspace*{-0.8cm}
\begin{tabular}{
ccccccccc%@{\extracolsep{3pt}}c@{\extracolsep{3pt}}c@{\extracolsep{3pt}}c@{\extracolsep{3pt}}l@{\extracolsep{3pt}}c@{\extracolsep{3pt}}c@{\extracolsep{3pt}}c@{\extracolsep{3pt}}c@{\extracolsep{3pt}}c@{\extracolsep{3pt}}c %
}
\cline{2-9}\noalign{\smallskip}\multirow{1}{*}{}&\multicolumn{4}{c}{Disregarding correlation}&\multicolumn{4}{c}{Considering correlation}\\
\cline{2-9}\noalign{\smallskip}\multirow{1}{*}{}&\multicolumn{4}{c}{Confidence level}&\multicolumn{4}{c}{Confidence level}\\
\hline\noalign{\smallskip}
$\beta$ & \multirow{1}{*}{0.50}  & \multirow{1}{*}{0.80} & \multirow{1}{*}{0.90} & \multirow{1}{*}{0.99} &\multirow{1}{*}{0.50}  & \multirow{1}{*}{0.80} & \multirow{1}{*}{0.90} & \multirow{1}{*}{0.99}\\
\hline\noalign{\smallskip}
\multirow{1}{*}{2.32}&4945.59&7802.42&10587.64&19868.91 &4856.48&  4926.56&5319.45&11684.41\\
\hline\noalign{\smallskip}
\end{tabular}
\color{black}
\label{2383oos}
\end{table}

\color{black}

\section{Conclusion}\label{Conclusions}

The consideration of some relevant information about uncertainty characterization in robust models for transmission network expansion planning is not explored yet in the literature. In this paper, the correlation between uncertain parameters has been explicitly considered to provide a least-cost expansion plan based on two-stage adaptive robust optimization. To that end, previously used cardinality and polyhedral uncertainty sets have been replaced with an ellipsoidal uncertainty set relying on the variance-covariance matrix of uncertain nodal net injections.
\reviewcol{The second stage of the proposed robust counterpart features a relevant analogy with the mathematical programming problem associated with the First-Order-Second-Moment method from structural reliability.}
\color{black}Based on this analogy, the resulting trilevel program is solved by the combined use of a column-and-constraint generation algorithm and a decomposition technique successfully applied in structural reliability. The proposed method is suitable for real-life large-scale systems. Moreover, such an analogy is useful for validation purposes and to provide the planner with guidelines on the choice of the conservativeness parameter.

From numerical results, the following conclusions are drawn:
\begin{enumerate}
  \item The proposed approach is able to capture the impact of correlation on power system planning.
  \item Incorporating correlation is relevant in terms of investment plans, thereby confirming the findings of \cite{BellWFH:15}.
  \item Expansion plans neglecting correlation are optimistic and underestimate load shedding.
  \item The higher the conservativeness parameter $\beta$ is, the higher the total cost is.
  \item Solutions considering correlation yield lower $\Phi(\beta)$-quantiles of sampled operating costs than those obtained when such a modeling aspect is ignored.
  \item Considering correlation usually results in higher investment costs than those obtained disregarding correlation. In those cases, the probability distribution of sampled operating costs is shifted to the left with respect to that when correlation is ignored.
 \item In case investment plans considering correlation are cheaper than those disregarding correlation, only the probability distribution of sampled operating costs below the desired confidence level $\Phi(\beta)$ is shifted to the left. This result comes at the expense of increasing the operating costs for higher quantiles.

\end{enumerate}

The proposed approach features three main limitations, namely 1) the model only uses information about the first two moments of the probability distribution of the uncertain parameters, 2) an approximation of the confidence level is provided, and 3) the lack of guaranteed convergence to global optimality of the inner loop does not ensure the attainment of global optimality. \color{black}
\reviewcol{Further work will address such limitations along with the characterization of correlated uncertainty via polyhedral uncertainty sets such as those used for robust generation scheduling and dispatch. Another interesting avenue of research is the improvement of the accuracy of the approximation provided by the First-Order-Second-Moment method for non-symmetrical mixed random variables such as the maximum wind power production.}
\appendix

This Appendix provides the detailed formulation of the lower-level problem (\ref{eq6})--(\ref{eq7}) presented in Section \ref{rtnep}-B. For further details, the interested reader is referred to \cite{MinguezGAA:17}.

\subsection{Nomenclature}

\subsubsection{Sets}
\begin{description}
[\IEEEsetlabelwidth{$\mbox{\it{ELLL}}$}\IEEEusemathlabelsep]

\item[$\Omega^{\rm D}$] Set of indexes $j$ of loads.

\item[$\Omega^{\rm D}_n$] Set of indexes $j$ of loads connected to bus $n$.
\vspace{0.1cm}

\item[$\Omega^{\rm G}$] Set of indexes $i$ of generating units.

\item[$\Omega^{\rm G}_n$] Set of indexes $i$ of generating units connected to bus $n$.

\item[$\Omega^{\rm L}$] Set of indexes $l$ of existing transmission lines.

\item[$\Omega^{\rm L^{\rm +}}$] Set of indexes $l$ of candidate transmission lines.

\item[$\Omega^{\rm N}$] Set of indexes $n$ of buses.

\end{description}

\subsubsection{Constants}
\begin{description}
[\IEEEsetlabelwidth{$\mbox{\it{ELLL}}$}\IEEEusemathlabelsep]

%%%\item[$\gamma_{j}$] Fraction of the demand of load $j$ that can be curtailed.

\item[$C^{\rm G}_{i}$] Production cost coefficient of unit $i$.
\vspace{0.1cm}

\item[$C^{\rm U}_{j}$] Load-shedding cost coefficient of load $j$.
\vspace{0.1cm}

\item[$fr(l)$] Sending or origin bus of line $l$.
\vspace{0.1cm}

\item[$\overline{P}^{\rm L}_{l}$] Power flow capacity of line $l$.

\item[$to(l)$] Receiving or destination bus of line $l$.

\item[$x_l$] Reactance of line $l$.

\end{description}

%\vspace{-0.2cm}

\subsubsection{Upper-Level Variables}

\begin{description}
[\IEEEsetlabelwidth{$\mbox{\it{ELLL}}$}\IEEEusemathlabelsep]

\item[$v_l$] Construction status of candidate line $l$ which is equal to 1 if the line is built, being 0 otherwise.

\end{description}

\subsubsection{Middle-Level Variables}

\begin{description}
[\IEEEsetlabelwidth{$\mbox{\it{ELLL}}$}\IEEEusemathlabelsep]

\item[$p^{\rm D}_{j}$] Demand of load $j$.
\vspace{0.1cm}

\item[$\overline p^{\rm G}_{i}$] Generation capacity of unit $i$.

\end{description}

\subsubsection{Lower-Level Variables}

\begin{description}
[\IEEEsetlabelwidth{$\mbox{\it{ELLL}}$}\IEEEusemathlabelsep]

\item[$\theta_n$] Phase angle at bus $n$.

\item[$p^{\rm G}_{i}$] Power output of unit $i$.
\vspace{0.1cm}

\item[$p^{\rm L}_{l}$] Power flow through line $l$.
\vspace{0.1cm}

\item[$p^{\rm U}_{j}$] Unserved demand of load $j$.
\vspace{0.1cm}

\end{description}

\subsection{Problem Formulation}

Based on \cite{MinguezGAA:17}, the lower-level problem (\ref{eq6})--(\ref{eq7}) is formulated as follows:

\begin{align}
%%%&\underset{\theta_{n},p^{\rm G}_{i},p^{\rm L}_{l},p^{\rm U}_{j}}{\text{Minimize}}  \quad  \sum_{i\in \Omega^{\rm G}}C^{\rm G}_{i}p^{\rm G}_{i}+\sum_{j\in \Omega^{\rm D}}C^{\rm U}_{j}p^{\rm U}_{j}- \epsilon^{\rm {G}} \!\!\!\sum_{\forall i \in {\Omega^{\rm G}}} \! \! \!\min{(0,\overline{p}^{\rm G}_{i})}\label{of3}\\
&\underset{\theta_{n},p^{\rm G}_{i},p^{\rm L}_{l},p^{\rm U}_{j}}{\text{Minimize}}  \quad  \sum_{i\in \Omega^{\rm G}}C^{\rm G}_{i}p^{\rm G}_{i}+\sum_{j\in \Omega^{\rm D}}C^{\rm U}_{j}p^{\rm U}_{j}\label{of3}\\
&\text{subject to:}\notag\\
&\sum_{i \in \Omega_n^{\rm G}}{p^{\rm G}_{i}}+\sum_{j \in {\Omega_n^{\rm D}}}{p^{\rm U}_{j}}+\sum_{l \in (\Omega^{\rm L}    \cup  \Omega^{\rm L^{\rm +}} )| {to}(l)=n}{p^{\rm L}_{l}} \nonumber\\
&- \sum_{l \in (\Omega^{\rm L}    \cup  \Omega^{\rm L^{\rm +}} )| {fr}(l)=n}{p^{\rm L}_{l}}= \sum_{j \in {\Omega_n^{\rm D}}}{p_{j}^{\rm D}};\forall n \in \Omega^{\rm N}\label{balance}\\
&p^{\rm L}_{l}=\frac{1}{x_l}(\theta_{fr(l)}-\theta_{to(l)}); \forall l \in \Omega^{\rm L}\label{flow_existing}\\
&p^{\rm L}_{l}=\frac{v_{l}}{x_l}(\theta_{fr(l)}-\theta_{to(l)}); \forall l \in \Omega^{\rm L^{\rm +}}\label{flow}\\
&-\overline{P}^{\rm L}_{l}\leq p^{\rm L}_{l}\leq \overline{P}^{\rm L}_{l}; \forall l \in (\Omega^{\rm L}    \cup  \Omega^{\rm L^{\rm +}} )\label{flow_lim}
\end{align}
\begin{align}
&0 \leq p^{\rm G}_{i} \leq \overline{p}^{\rm G}_{i};\forall i \in {\Omega^{\rm G}}\label{gen_lim}\\
%%%&0 \leq p^{\rm G}_{i} \leq \max{(0,\overline{p}^{\rm G}_{i})};\forall i \in {\Omega^{\rm G}}\label{gen_lim}\\
%\end{align}
%\begin{align}
&0 \leq p^{\rm U}_{j} \leq p^{\rm D}_{j};\forall j \in {\Omega^{\rm D}}.\label{shed_lim}%\\
\end{align}

Problem (\ref{of3})--(\ref{shed_lim}) is driven by the minimization of the operating cost (\ref{of3}) for the values of $v_l$ determined in the upper level and the values of $p^{\rm D}_{j}$ and $\overline p^{\rm G}_{i}$ identified in the middle level. Expressions (\ref{balance})--(\ref{flow_lim}) model the effect of the network including nodal power balances  (\ref{balance}), line flows through existing lines (\ref{flow_existing}), line flows through candidate lines (\ref{flow}), and line flow limits (\ref{flow_lim}). Constraints (\ref{gen_lim}) set the generation limits. Finally, constraints (\ref{shed_lim}) impose bounds on load shedding.

%%%%%%%%%%%%%%%%%%%%%%%%%%%%%%%%%%%%%%%%%%%%%%%%%%%%%%%%%%%%%%%%%%%%%%%%%%%%%%%%%%%%%%%%%%%%
%                                                           Thebibliography
%%%%%%%%%%%%%%%%%%%%%%%%%%%%%%%%%%%%%%%%%%%%%%%%%%%%%%%%%%%%%%%%%%%%%%%%%%%%%%%%%%%%%%%%%%%%
%%%\bibliographystyle{IEEEtran}% basic style, author-year citations
%%%\bibliography{Biblio}

\begin{IEEEbiographynophoto}{Cristina Rold\'an} received the B.Eng. degree in electrical engineering and the M.Eng. degree in industrial engineering from the Universidad de Castilla-La Mancha, Ciudad Real, Spain, in 2014 and 2016, respectively.

She is currently pursuing the Ph.D. degree in electrical engineering at the Universidad de Castilla-La Mancha. Her research interests include operations, planning, and economics of power systems.
\end{IEEEbiographynophoto}

\vspace{-1 cm}

\begin{IEEEbiographynophoto}{Roberto M\'{\i}nguez} received the Civil Engineer degree and the Ph.D.
degree from the Universidad de Cantabria, Santander,
Spain, in 2000 and 2003, respectively.

He is currently an independent consultant. His research interests
include reliability engineering, sensitivity analysis, numerical methods, and optimization.
\end{IEEEbiographynophoto}

\vspace{-1 cm}

\begin{IEEEbiographynophoto}{Raquel Garc\'ia-Bertrand}
(S'02--M'06--SM'12) received the Ingeniera Industrial degree and the Ph.D.
degree from the Universidad de Castilla-La Mancha, Ciudad Real,
Spain, in 2001 and 2005, respectively.

She is currently an Associate Professor of electrical engineering at
the Universidad de Castilla-La Mancha. Her research interests
include operations, planning, and economics of electric
energy systems, as well as optimization and decomposition
techniques.
\end{IEEEbiographynophoto}

\vspace{-1 cm}

\begin{IEEEbiographynophoto}{Jos\'e M. Arroyo}
(S'96--M'01--SM'06) received the Ingeniero Industrial degree from the
Universidad de M\'alaga, M\'alaga, Spain, in 1995, and the Ph.D. degree
in power systems operations planning from the Universidad de
Castilla-La Mancha, Ciudad Real, Spain, in 2000.

From June 2003 through July 2004 he held a Richard H. Tomlinson Postdoctoral Fellowship at the Department of Electrical and Computer Engineering of McGill University, Montreal, QC, Canada. He is
currently a Full Professor of electrical engineering at the
Universidad de Castilla-La Mancha. His research interests include
operations, planning, and economics of power systems, as well as
optimization.
\end{IEEEbiographynophoto}

\end{document}